%% file: 00_abstarct.tex
\documentclass[sigconf]{acmart}
\usepackage{hyperref}
\usepackage{amsmath}
\usepackage{graphicx}
\usepackage{textcomp}
\usepackage{xcolor}
\usepackage{listings}
\usepackage[ruled, linesnumbered]{algorithm2e}
\usepackage{lipsum}
\usepackage{multirow}
\usepackage{booktabs}
\usepackage{threeparttable}
\usepackage{url}
\usepackage{fancyhdr}
\usepackage{indentfirst}
\usepackage{resizegather}
\usepackage{colortbl}
\usepackage{dblfloatfix}
\usepackage{makecell}
\usepackage{array}
\usepackage{enumitem}
\usepackage{hhline}
\usepackage{float}
\usepackage[binary-units]{siunitx}
\usepackage[export]{adjustbox}
\usepackage{subcaption}
\usepackage[hyphens]{xurl}
\usepackage{array}
\usepackage{soul}
\usepackage{tcolorbox}
\usepackage{xspace}
\usepackage{tikz}

\AtBeginDocument{%
  }

\copyrightyear{2025}
\acmYear{2025}
\setcopyright{cc}
\setcctype{by}
\acmConference[SoCC '25]{ACM Symposium on Cloud Computing}{November 19--21, 2025}{Online, USA}
\acmBooktitle{ACM Symposium on Cloud Computing (SoCC '25), November 19--21, 2025, Online, USA}\acmDOI{10.1145/3772052.3772215}
\acmISBN{979-8-4007-2276-9/2025/11}

\newcommand*\circled[1]{\tikz[baseline=(char.base)]{
    \node[shape=circle,draw,inner sep=1pt] (char) {#1};}}

\newcommand{\PagedAttention}{\textit{PagedAttention}\xspace}

\newcommand{\Prompt}{\textit{prefill}\xspace}
\newcommand{\TokenGen}{\textit{decode}\xspace}

\newcommand{\LLMPipe}{\textit{Oneiros}\xspace}
\newcommand{\Engine}{\textit{Remapping Controller}\xspace}

\definecolor{LightBlue}{rgb}{0.83, 0.91, 1}
\definecolor{Red}{rgb}{1, 0, 0}

\begin{document}

\title{Oneiros: KV Cache Optimization through Parameter Remapping for Multi-tenant LLM Serving}

\author{Ruihao Li$^{\text{1}, *}$\enskip Shagnik Pal$^{\text{1}, *}$\enskip Vineeth Narayan Pullu$^{\text{1}}$\enskip Prasoon Sinha$^{\text{1}}$\enskip 
\\ Jeeho Ryoo$^{\text{2}}$\enskip Lizy K. John$^{\text{1}}$\enskip Neeraja J. Yadwadkar$^{\text{1}}$}
\affiliation{\vspace{1mm} $^{\text{1}}$The University of Texas at Austin \enskip $^{\text{2}}$Fairleigh Dickinson University \country{}}

\renewcommand{\shortauthors}{R. Li, S. Pal, et al.}

\begin{abstract}
KV cache accelerates LLM inference by avoiding redundant computation, at the expense of memory. 
To support larger KV caches, prior work extends GPU memory with CPU memory via CPU-offloading. 
This involves swapping KV cache between GPU and CPU memory. 
However, because the cache updates dynamically, such swapping incurs high CPU memory traffic.
We make a key observation that model parameters remain constant during runtime, unlike the dynamically updated KV cache. 
Building on this, we introduce \LLMPipe, which avoids KV cache swapping by remapping, and thereby repurposing, the memory allocated to model parameters for KV cache. 
This parameter remapping is especially beneficial in multi-tenant environments, where the memory used for the parameters of the inactive models can be more aggressively reclaimed.
Exploiting the high CPU-GPU bandwidth offered by the modern hardware, such as the NVIDIA Grace Hopper Superchip, we show that \LLMPipe significantly outperforms state-of-the-art solutions, achieving a reduction of 44.8\%-82.5\% in tail time-between-token latency, 20.7\%-99.3\% in tail time-to-first-token latency, and 6.6\%-86.7\% higher throughput compared to vLLM.
Source code of \LLMPipe is available at \url{https://github.com/UT-SysML/Oneiros/}.
\end{abstract}



\maketitle

\input{content}

\bibliographystyle{ACM-Reference-Format}
\bibliography{refs}
\end{document}

%% file: content.tex
\section{Introduction}
\label{section:introduction}
\input{01_introduction}

\vspace{-2mm}
\section{Background}
\label{section:background}
\input{02_background}

\vspace{-2mm}
\section{Motivation}
\label{section:motivation}
\input{03_motivation}

\vspace{-3mm}
\section{\LLMPipe Overview}
\label{section:design}
\input{04_design}

\vspace{-2mm}
\section{Remapping Controller}
\label{section:engine}
\input{05_00_engine}

\vspace{-2mm}
\section{Implementation}
\label{section:implementation}
\input{06_implementation}

\vspace{-2mm}
\section{Evaluation}
\label{section:evaluation}
\input{07_00_evaluation}

\section{Discussion and Related Work}
\label{section:discussion}
\input{08_discussion}

\vspace{-3mm}
\section{Conclusion}
\label{section:conclusion}
\vspace{-1mm}
\input{09_conclusion}

%% file: 01_introduction.tex
\begin{figure}[t]
        \begin{minipage}[t]{0.52\columnwidth}
		\centering	
        \includegraphics[width=\columnwidth,trim = 6mm 6mm 6mm 6mm, page=1, clip=true]{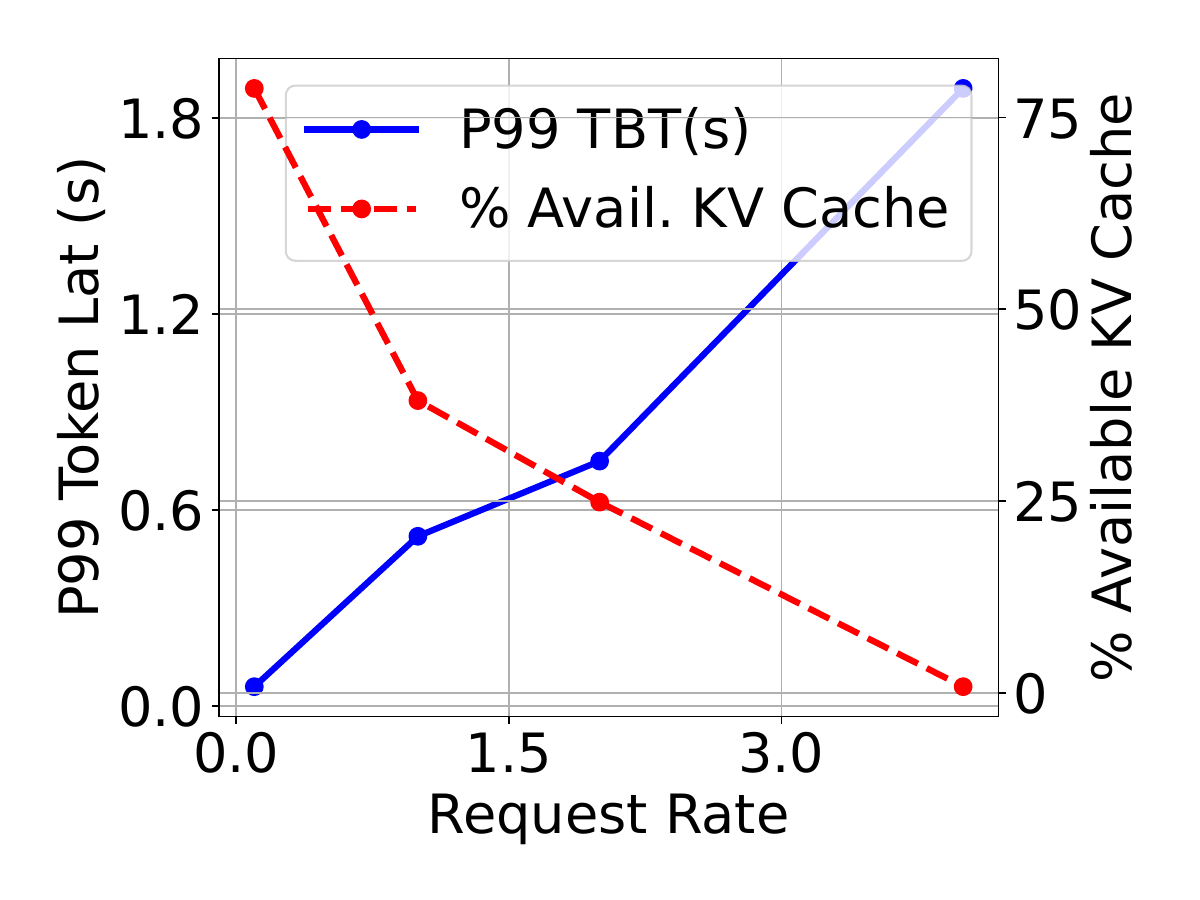}
        \vspace{-6mm}
		\subcaption{Existing LLM systems.}
		\label{fig_rps_vs_kv}
	\end{minipage}  
	\begin{minipage}[t]{0.46\columnwidth}
		\centering
		\includegraphics[width=\columnwidth,trim = 6mm 6mm 6mm 6mm, page=1, clip=true]{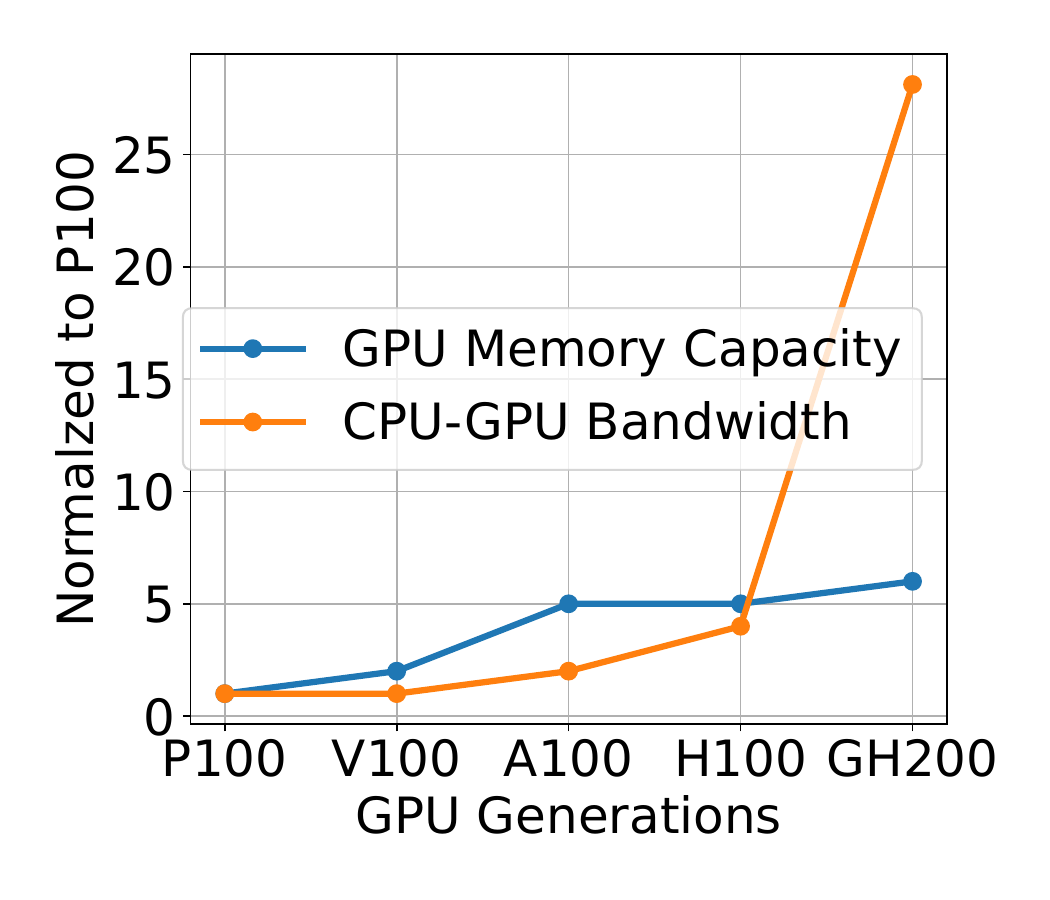}
        \vspace{-6mm}
		\subcaption{GPU Trends.}
		\label{fig_gpu_trends}
	\end{minipage}
    \vspace{-4mm}
	\caption{(a) Existing LLM systems perform recomputation when KV cache is exhausted, leading to a significant increase in latency under high request rates (OPT-13b with ShareGPT). (b) CPU-GPU bandwidth in Nvidia GPUs has increased significantly since GH200, bringing CPU-offloading opportunities. 
    }
    \vspace{-2mm}
	\label{fig_intro}
\end{figure}

{\let\thefootnote\relax\footnote{{$^*$Equal contribution.}}\addtocounter{footnote}{-1}}
The memory demands of Large Language Models (LLMs) are growing at a much faster rate than GPU memory capacity, making memory a major bottleneck for efficient inference serving~\cite{hanindhito2024bandwidth, kwon2023efficient, touvron2023llama2, dubey2024llama, A100, H100, GH200}.
KV cache~\cite{pope2023efficiently, kwon2023efficient} was introduced as a key mechanism to reduce inference latency: it stores intermediate results from previously generated tokens to avoid recomputing tensor values. 
To harness the impact of KV cache on the performance of LLM inference serving, numerous memory management techniques are proposed~\cite{pope2023efficiently, lee2024infinigen, liu2024cachegen, prabhu2024vattention, jin2024compute, adnan2024keyformer}, including paging-based mechanisms for runtime KV cache management~\cite{kwon2023efficient}.
However, despite these optimizations, when the KV cache size hits the limit of the GPU memory, one must recompute key-value pairs, increasing serving latency~\cite{kwon2023efficient} (Figure~\ref{fig_rps_vs_kv}).
The KV-cache bottleneck increases further in multi-tenant environments where multiple LLMs share and contend over the GPU(s) resources~\cite{yu2025prism}. \par

Using CPU memory as an extension of the GPU memory (also called CPU-offloading) is one way to ease the pressure on GPU memory~\cite{peng2020capuchin, huang2020swapadvisor, rajbhandari2021zero, ren2021zero, lim2021zico, sheng2023flexgen, jiang2024neo, xu2024pie, naflexinfer}.
However, it comes at the cost of increased latency or reduced throughput, depending on the CPU-GPU transfer bandwidth. 
Prior work~\cite{jiang2024neo} explored offloading portions of attention computation to CPUs, allowing KV cache to be generated and stored in CPU memory rather than occupying GPU memory.
This allowed the CPU-GPU system to handle larger batch sizes; however, limited CPU performance in attention computation can become a bottleneck, potentially increasing token generation latency by about 10-20\% in certain scenarios~\cite{jiang2024neo}. \par

To make it practical, new advancements from the hardware community have dramatically increased the CPU-GPU data transfer bandwidth: the latest Grace Hopper superchip (GH200)~\cite{GH200} introduces a new NVLink-C2C interface between CPU and GPU, offering $7\times$ bandwidth of the H100~\cite{H100} and $14\times$ that of the A100~\cite{A100} (Figure~\ref{fig_gpu_trends}).
This increased CPU-GPU bandwidth opens new opportunities to design low-latency and high-throughput LLM inference serving systems. 
At the same time, with the availability of higher bandwidth, the limited parallelism of CPUs, rather than the bandwidth, may emerge as the new performance bottleneck (\S~\ref{section_compute_swap}). \par

\begin{figure}[t]
    \centerline{\includegraphics[width=\columnwidth, trim = 17mm 50mm 21mm 15mm, page=7, clip=true]{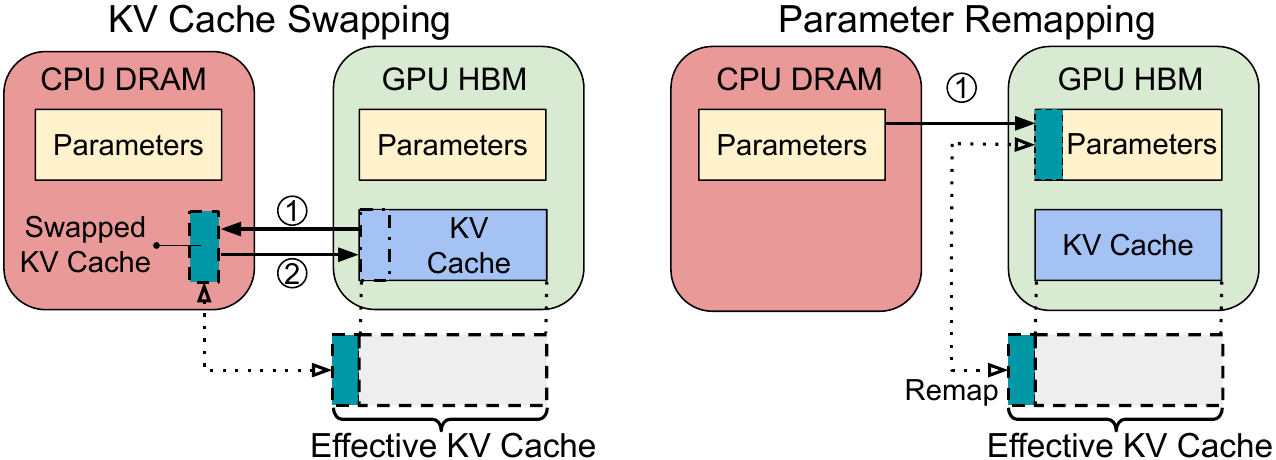}}
    \vspace{-3mm}
	\caption{
    Both KV cache swapping and parameter remapping expand the effective KV cache capacity by utilizing CPU memory as an extension of GPU memory.
    KV cache swapping incurs overhead due to frequent synchronization, as it requires backing up KV cache to CPU memory after each token generation once swap begins.
    Instead, parameter remapping is unidirectional with negligible synchronization overhead.
    }
    \vspace{-3mm}
	\label{fig_para_kv}
\end{figure}

Existing work, such as Pie~\cite{xu2024pie}, leverages this extra CPU-GPU bandwidth by offloading the KV cache of future layers to CPU memory and swapping back to GPU memory as required, thereby increasing the effective KV cache size (see Figure~\ref{fig_para_kv} (left side)). 
This allows for larger batches or longer sequences without triggering recomputation, thereby reducing inference serving latency and improving serving throughput.
However, the frequent updates to the KV cache in GPU memory require synchronization during KV cache swapping, introducing overheads with bidirectional data transfers.
As shown in Figure~\ref{fig_para_kv} (left side), the KV cache in GPU memory cannot be freed or overwritten (step 2) until the swap to CPU memory (step 1) is complete, since the KV cache is updated after each token generation iteration.
We empirically note that this dependency introduces frequent synchronization between CPU and GPU memory during KV cache swapping, leading to increased serving latency and reduced system throughput (see \S~\ref{section_kv_swap}). \par

Instead, we propose \emph{dynamic parameter remapping}\footnote{In this paper, swap refers to bidirectional data transfer to and from CPU/GPU memory, whereas remap denotes unidirectional data transfer from CPU to GPU memory.}, where we simply repurpose a portion of the GPU memory originally allocated for model parameters to support a larger KV cache, enabling the system to accommodate increased KV cache requirements.
Unlike KV cache swapping, parameter remapping introduces negligible synchronization overhead, as model parameters remain constant during inference. 
As shown in Figure~\ref{fig_para_kv} (right), it is a non-blocking, unidirectional data transfer that does not require offloading parameters back to the CPU. 
This immutability simplifies resource management, as the transfer is both deterministic and fixed in size. 
Notably, parameter remapping is particularly well-suited for multi-tenant LLM serving: it enables efficient and proactive reuse of memory allocated to inactive models\footnote{Prior work observed that models frequently enter idle periods without incoming requests~\cite{yu2025prism}. We refer to models in such intervals as inactive models.} by repurposing their parameter storage as KV cache for active models. \par

However, realizing a practical parameter remapping engine raises several open questions, including when to initiate and stop remapping, which models to target, how many layers to remap, and which specific layers to select.
For example, suboptimal layer selection can trigger excessive CPU-GPU data transfers, resulting in a 12.7\% reduction in throughput (\S~\ref{section_effective_layer}).
Failing to halt remapping in a timely manner can introduce unnecessary pipeline overhead~\cite{hestness2015gpu, xu2024pie}, negatively impacting system performance and increasing serving latency by up to 49\% (\S~\ref{section_ablation}).
The complexity of these decisions is further amplified by the dynamic nature of system load and the variability in request sequence lengths across inputs~\cite{yu2025prism, stojkovic2024dynamollm, patel2024splitwise}.
As a result, these decisions must be made adaptively at runtime, highlighting the need for a sophisticated memory management and data transfer engine within the LLM inference serving system (\S~\ref{section_characterization_time_model}). \par

In this paper, we introduce \LLMPipe, a \textit{Dynamic Remapping Engine} that automatically and elastically adjusts KV cache size at \textit{layer} granularity based on \textit{runtime} demands. 
Such flexibility allows \LLMPipe to be integrated into \textit{any LLM inference serving system with any schedulers} (\S~\ref{section:discussion}). 
We make the following key contributions:

\begin{itemize}
    \item We study several GPU memory management optimization strategies to motivate dynamic parameter remapping.
    \item We introduce \LLMPipe, a \textit{Dynamic Remapping Engine} for multi-tenant LLM inference serving systems that elastically adjusts the KV cache size at runtime by remapping parameter memory as KV cache memory.
    \item We create a layer eviction algorithm that leverages the circular execution characteristics of LLMs to select which layers to remap. 
    In addition to empirical evaluation, we provide a theoretical proof showing the optimality of our algorithm.
    \item We integrate \LLMPipe into vLLM to evaluate its effectiveness. Compared to the baseline vLLM, \LLMPipe increases throughput by 7.9\%-86.7\% and reduces tail latency by 20.7\%-99.3\%.
\end{itemize}

%% file: 02_background.tex
This section provides background on the state-of-the-art inference serving system  (\S~\ref{section_llm_serving_gpu_background}), highlighting that KV cache is the primary bottleneck in both scenarios (\S~\ref{section_challenge}).

\subsection{LLM Inference Serving}
\label{section_llm_serving_gpu_background}
\begin{figure}[t]
    \centerline{\includegraphics[width=0.9\columnwidth, trim = 28mm 34mm 61mm 13mm, page=2, clip=true]{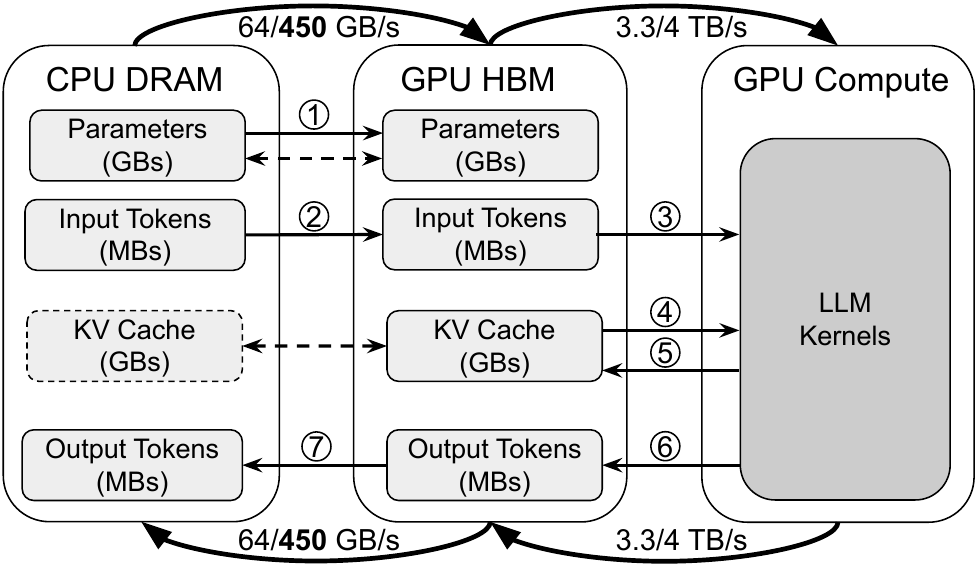}}
    \vspace{-4mm}
	\caption{LLM inference serving on GPUs. 
    GH200 features 450 GB/s CPU-GPU bandwidth, while H100 only has 64 GB/s. 
    Swapping KV cache/parameter between CPUs and GPUs during runtime can fully utilize the increased bandwidth.}
	\label{fig_llm_gpu_background}
    \vspace{-2mm}
\end{figure}

\noindent{\textbf{Inference Execution Workflow:}}
LLM inference serving consists of two phases: (a) the \Prompt phase, which processes the full input sequence in a single pass, and (b) the \TokenGen phase, which generates one output token at a time. 
Figure~\ref{fig_llm_gpu_background} illustrates the data placement and execution workflow for a typical LLM inference serving system~\cite{kwon2023efficient}. 
The two primary contributors to GPU memory are LLM model parameters and the KV Cache~\cite{pope2023efficiently, kwon2023efficient, agrawal2024taming, prabhu2024vattention}. 
\circled{1} Before the inference serving phase, the model parameters are transferred from CPU memory (DRAM) to GPU memory (HBM)~\cite{li2023alpaserve, kwon2023efficient, agrawal2024taming, prabhu2024vattention, fu2024serverlessllm}.
After loading parameters, \circled{2} input tokens are then loaded into GPU memory, where \circled{3} tokens are transferred to GPU compute units (where LLM kernel executes) for processing during the \Prompt phase.
During the \TokenGen phase, the KV cache is generated and expands with the sequence length.
Throughout each iteration, the KV cache is transferred between \circled{4} LLM kernels and \circled{5} GPU memory.
When the maximum sequence length is reached or an end-of-sequence (EOS) token is generated, the output tokens produced by the GPU compute units are first written to GPU memory \circled{6}, and then transferred back to CPU memory \circled{7}.

\noindent{\textbf{KV Cache:}} 
The KV cache stores intermediate results from previously generated tokens during the \TokenGen phase, eliminating the need to recompute previously generated tensor values.
Its size scales linearly with both the batch size and request sequence length.

\noindent{\textbf{Performance Metrics:}}
The performance of an LLM inference serving system is typically characterized by two latency metrics --\textit{TTFT} (time-to-first-token) and \textit{TBT} (time-between-tokens) -- and by \textit{throughput}.
TTFT measures the delay before the first token is generated, indicating system responsiveness, while TBT captures the interval between consecutive tokens, reflecting response smoothness.
Throughput quantifies how many requests or tokens the system can process per unit time. \par

\vspace{-2mm}
\subsection{Memory Bottlenecks in LLM  Serving }
\label{section_challenge}
The memory demands of LLMs have increased at a much faster rate than GPU memory capacity, making memory a critical bottleneck in LLM inference serving. 
As model sizes grow, they require not only greater computational resources but also significantly more memory to accommodate both model parameters and KV cache~\cite{kwon2023efficient, agrawal2024taming}. 
To address this challenge, recent works focused on \textit{efficiently managing KV cache}, as its size grows linearly with the sequence length and batch size, which are unknown prior to runtime~\cite{pope2023efficiently, kwon2023efficient, lee2024infinigen, liu2024cachegen, prabhu2024vattention, jin2024compute, adnan2024keyformer}. 
For example, vLLM employs \PagedAttention~\cite{kwon2023efficient}, which leverages a paging mechanism to manage KV cache pages during runtime.
However, when the KV cache exceeds the available GPU memory, recomputation becomes necessary. 
This scenario is difficult to anticipate, as the sequence length is unknown before runtime and request arrival patterns vary in production environments~\cite{stojkovic2024smartoclock, patel2024splitwise, stojkovic2024dynamollm}.
This unpredictability also affects multi-tenant LLM inference systems.
Regardless of the scheduling policy, systems using \PagedAttention face the same challenge as single-model serving: runtime KV cache usage is inherently unpredictable.
Recomputation can significantly degrade the performance of the affected sequence, especially by increasing the tail latency of the overall serving system.
To mitigate this, a mechanism is desired to prevent recomputation when KV cache capacity is exceeded, thereby reducing tail latency in LLM inference serving. \par

%% file: 03_motivation.tex
This section presents three approaches to expanding KV cache capacity, enabled by the significantly higher CPU-GPU bandwidth of modern hardware architectures like GH200. 
\begin{itemize}
    \item Approach 1: offloading computation to CPUs~\cite{jiang2024neo}, where a portion of the attention operations is offloaded to CPUs, allowing the KV cache to be generated and stored in CPU memory instead of consuming GPU memory. 
    As a result, the combined CPU-GPU system can accommodate larger batch sizes, leading to increased throughput (\S~\ref{section_compute_swap}). 
    \item Approach 2: KV cache swapping~\cite{xu2024pie}, which swaps out portions of the KV cache to CPU memory when not immediately needed, freeing up GPU memory for active KV cache usage.
    This allows the system to accommodate larger batch sizes within the available GPU memory (\S~\ref{section_kv_swap}).
    \item Approach 3 (proposed): \emph{parameter remapping}, which dynamically repurposes model parameter memory to automatically and elastically expand KV cache capacity at runtime, enabling support for larger batches or longer sequences (\S~\ref{section_why_parameter_remapping}). 
\end{itemize}

KV cache swapping (Approach 2) and parameter remapping (Approach 3) better utilize the available bandwidth without incurring substantial performance overhead and are more effective than offloading computation to CPUs (Approach 1), which is primarily suited to legacy systems with limited bandwidth. 
We further compare KV cache swapping (Approach 2) with parameter remapping (Approach 3) and show that parameter remapping has the potential to offer even better performance, particularly in multi-tenant LLM inference scenarios.
We conclude with a quantitative analysis of computation and parameter loading times, highlighting challenges and the importance of efficient parameter remapping (\S~\ref{section_characterization_time_model}).
\par

\vspace{-2mm}
\subsection{Limitations of CPU Offloading}
\label{section_compute_swap}
\begin{figure}[t]
        \begin{minipage}[t]{0.49\columnwidth}
		\centering	
        \includegraphics[width=\columnwidth,trim = 6mm 8mm 5mm 5mm, page=1, clip=true]{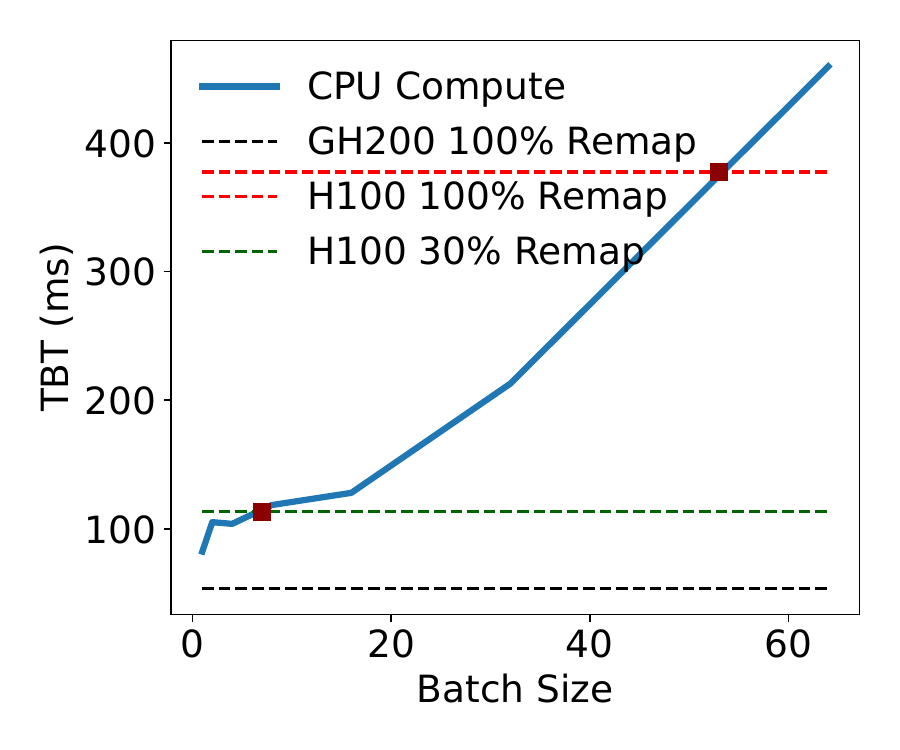}
        \vspace{-5mm}
		\subcaption{CPU Compute.}
		\label{fig:charac-cpu-times}
	\end{minipage}
	\begin{minipage}[t]{0.49\columnwidth}
		\centering
		\includegraphics[width=\columnwidth,trim = 6mm 8mm 5mm 6mm, page=1, clip=true]{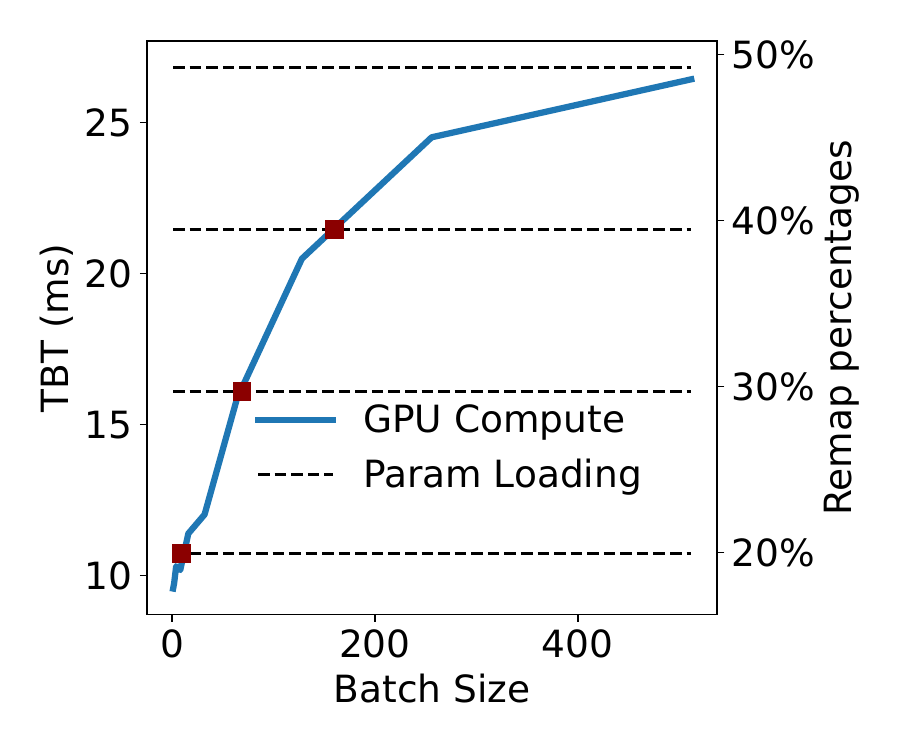}
        \vspace{-5mm}
		\subcaption{GPU Compute.}
		\label{fig:charac-gpu-times}
	\end{minipage} 
    \vspace{-4mm}
	\caption{(a) Computation time of different batch sizes when offloading to CPUs. Parameter remapping becomes a more promising approach when CPU-GPU bandwidth is sufficient. (b) GPU computation time fluctuates with batch size based on the load in an online serving system, requiring the parameter remapping percentage to be adjusted dynamically.
     }
	\label{fig_character_gpu_cpu}
    \vspace{-4mm}
\end{figure}

Prior works have demonstrated that offloading part of the attention computation and KV cache states from GPUs to CPUs can effectively increase the batch size and thus inference throughput in LLM inference serving systems~\cite{jiang2024neo}.
Although offloading computation to CPUs has shown promising performance on earlier hardware platforms, we find it is less efficient on advanced systems such as the GH200 superchip, where limited CPU parallelism, rather than CPU-GPU bandwidth, is the bottleneck.
Figure~\ref{fig:charac-cpu-times} presents the CPU computation time for processing OPT-13b across varying batch sizes serving requests from the ShareGPT dataset~\cite{sharedgpt}. 
We compare two approaches: (a) offloading computations to CPUs, and (b) continuing GPU computation while remapping parameters to expand KV cache capacity to support large batch sizes.
Parameter remapping is preferable when the parameter loading time does not exceed the CPU computation time. 
In GH200 systems, high CPU-GPU bandwidth but limited CPU parallelism make parameter loading significantly faster than CPU processing, reinforcing the suitability of parameter remapping.
In contrast, on an H100 system, offloading computation to CPUs may be viable, even when compared to a low remapping percentage (the fraction of memory used by model parameters that are remapped as added capacity for KV cache), such as 30\%. 
Therefore, on systems with high CPU-GPU bandwidth, remapping parameters offers a more promising solution than offloading computation to CPUs. \par

\vspace{-2mm}
\subsection{Limitations of Swapping KV Cache}
\label{section_kv_swap}
With emerging hardware offering higher CPU-GPU bandwidth, KV cache for non-executing (future) layers can be temporarily swapped into CPU memory and brought back to GPU memory when needed. 
This approach enables larger batch sizes and enhances overall system performance, as demonstrated by Pie~\cite{xu2024pie}.
However, KV cache swapping presents significant challenges, primarily because the KV cache must be updated after every token generation~\cite{yu2022orca, kwon2023efficient, agrawal2024taming}.
This results in frequent synchronization between CPU and GPU memory, introducing overhead that affects serving latency.
In addition to synchronization overhead, the bidirectional nature of data transfers can reduce effective CPU-GPU bandwidth, preventing the system from achieving an ideal pipeline of data transfer and computation.
Our measurements of CPU host memory (LPDDR5X\footnote{The trend observed in our measurements also holds for other memory technologies (e.g., DDR4, DDR5) beyond LPDDR5X~\cite{esmaili2024mess}.}) bandwidth show that unidirectional (read-only of CPU host memory) CPU-to-GPU transfers can reach $\sim$427 GB/s. 
However, when the read-to-write ratio is 1:1, bandwidth drops to $\sim$366 GB/s, resulting in a $\sim$15\% reduction.
This bandwidth degradation can further impact serving throughput. \par

\subsection{Our Proposal: Parameter Remapping}
\label{section_why_parameter_remapping}
\begin{figure}[t]
    \begin{minipage}[t]{\columnwidth}
		\centering	\includegraphics[width=\columnwidth,trim = 29mm 43mm 10mm 40mm, page=5, clip=true]{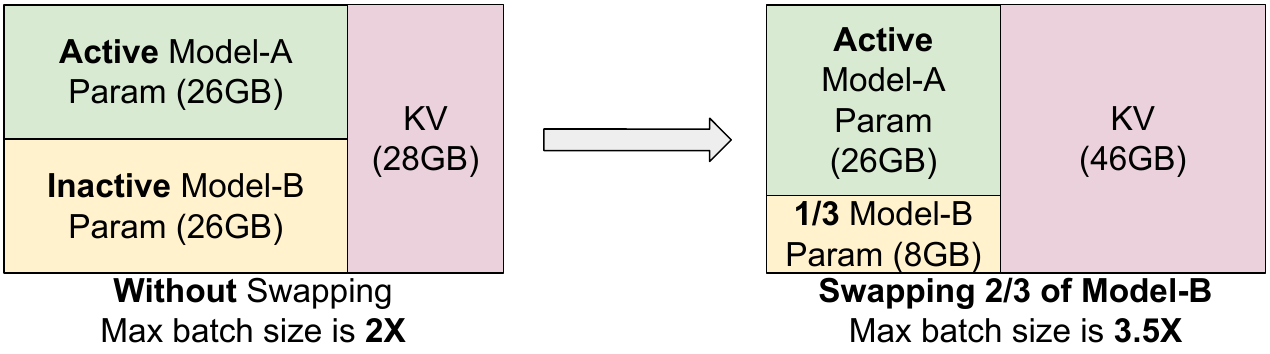}
        \vspace{-6mm}
		\subcaption{Temporal sharing (remap 2/3 param of inactive Model-B).}
        \label{fig_without_llm_pipe_1}
        \vspace{1mm}
	\end{minipage}
	\begin{minipage}[t]{\columnwidth}
		\centering
		\includegraphics[width=\columnwidth,trim = 29mm 36mm 10mm 40mm, page=4, clip=true]{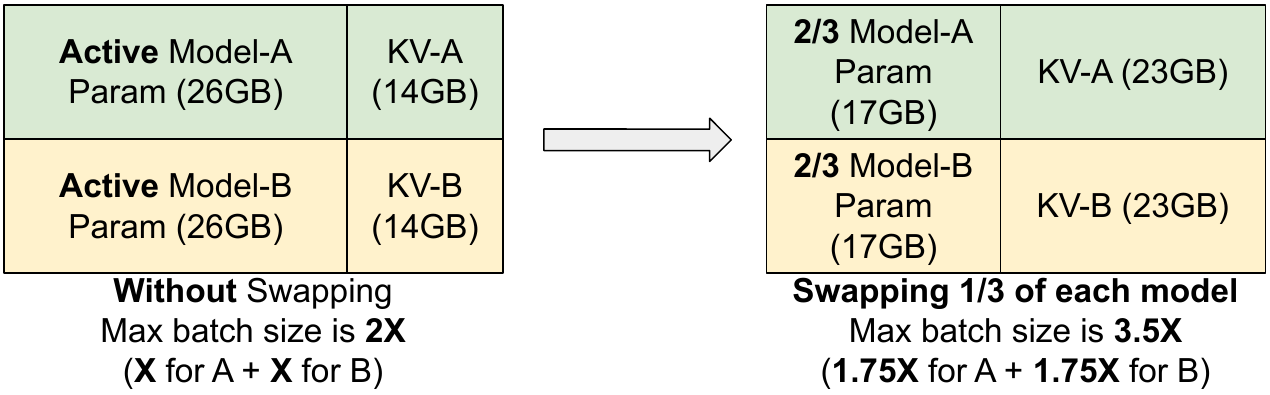}
        \vspace{-6mm}
		\subcaption{Spatial sharing (remap 1/3 param of each model).}
	\label{fig_without_llm_pipe_2}
	\end{minipage}
    \vspace{-4mm}
	\caption{Remapping parameters for KV cache allows for larger batches when deploying two OPT-13b models on H100, in both (a) temporal sharing and (b) spatial sharing scenarios. }
    \vspace{-4mm}
	\label{fig_without_llm_pipe}
\end{figure}

\textit{We introduce Parameter Remapping, where, instead of swapping KV cache, we propose to dynamically remap the space used for model parameters to automatically and elastically expand the KV cache size at runtime.}
Dynamic parameter remapping reclaims GPU memory by repurposing a portion of the model parameters for KV cache usage during runtime, enabling online inference serving systems to handle peak loads when GPU memory becomes insufficient for KV cache.
As shown in the two examples in Figure~\ref{fig_without_llm_pipe}, deploying two OPT-13b models on an H100 GPU using either temporal (Figure~\ref{fig_without_llm_pipe_1}) or spatial (Figure~\ref{fig_without_llm_pipe_2}) sharing one GPU consumes approximately 52GB of the available 80GB memory for model parameters~\cite{kwon2023efficient}, leaving only about 28GB for KV cache.
By remapping one-third of the model parameters as KV Cache -- either $2/3$ from the inactive model in temporal sharing or $1/3$ from each model in spatial sharing -- the maximum batch size can be increased by up to $1.75\times$, resulting in a corresponding $1.75\times$ improvement in system throughput.
Unlike KV cache swapping, parameter remapping involves non-blocking, unidirectional data transfers and does not require backing up model parameters from GPU to CPU, as parameters remain unchanged during inference.
As a result, synchronization overhead is significantly reduced.
Additionally, parameter remapping is particularly effective in multi-tenant LLM inference scenarios, such as multi-agent workflows~\cite{talebirad2023multi, chan2023chateval, li2024more, han2024llm, kim2025cost} and production deployments where many models remain idle for long periods of time~\cite{yu2025prism}.
In such settings, memory allocated to inactive models can be reclaimed for active ones, a flexibility that KV cache swapping lacks, as inactive models may have no KV cache to be swapped. \par

\subsection{Challenges in Parameter Remapping}
\label{section_characterization_time_model}
While parameter remapping can improve performance, it introduces potential overhead and requires careful design.
To avoid impacting inference latency and throughput, parameter loading must ideally overlap with GPU computation, ensuring that the parameters of each layer are fully loaded before its execution begins. 
For example, while the GPU  executes layer 1, layer 2's parameters can be concurrently loaded from CPU to GPU memory.
The ideal case is when layer 2 finishes loading before layer 1 finishes execution, allowing GPU memory to hold only two layers at a time~\cite{sheng2023flexgen, deepspeed}, unlike approaches that preload the entire model into GPU memory.
Despite GH200's higher CPU-GPU bandwidth, it remains insufficient for full on-demand remapping, since layer execution outpaces parameter transfer.
As a result, only a fraction of the parameters, defined by the \textit{remapping percentage}, can be dynamically transferred at runtime, while the rest must be preloaded into GPU memory before the inference serving starts.
In this section, we examine the challenges involved in making these design decisions.
\par

\noindent\textbf{When to trigger remapping?}
While an ideal pipeline can fully overlap GPU computation with parameter loading from CPU to GPU memory, the parameter remapping engine needs to initiate parameter remapping only when necessary, as even a perfectly pipelined approach introduces inherent overhead~\cite{hestness2015gpu}. 
Likewise, it is important to halt parameter remapping once it is no longer needed.
Delaying this decision can increase the latency of the serving system by up to 49\% (\S~\ref{section_ablation}).
These decisions require dedicated mechanisms and careful design for online monitoring of the LLM inference serving system. \par

\noindent\textbf{Which models to remap?}
In multi-tenant LLM inference serving systems, not all models are active simultaneously~\cite{patke2024queue, yu2025prism}, which presents an opportunity for the parameter remapping engine to prioritize remapping parameters of inactive models.
However, this approach requires careful design, as overly aggressive offloading may lead to starvation -- especially given the unpredictable nature of request arrival patterns in online serving environments~\cite{stojkovic2024dynamollm}.
An inefficient design is likely to prevent certain models from responding promptly; for example, a suboptimal model selection can lead to a 22.0\% increase in serving tail latency (\S~\ref{section_ablation}). \par

\noindent{\textbf{How many layers to remap?}} 
Determining the optimal remapping percentage is challenging due to the dynamic nature of request arrival rates in real-world online inference serving~\cite{stojkovic2024smartoclock, patel2024splitwise, stojkovic2024dynamollm}. 
The fluctuations in request arrival rates can affect the GPU computation time, which in turn affects the percentage of parameters that can be transferred without performance degradation.
We illustrate the need for dynamic adjustment of the remapping percentage using an example of serving OPT-13b with varying batch sizes on the ShareGPT dataset~\cite{sharedgpt}. 
Figure~\ref{fig:charac-gpu-times} shows how the GPU computation time for generating one token in the \TokenGen phase varies with batch size, while parameter load time -- determined solely by the remapping percentage -- remains constant. 
The intersection points of these curves represent configurations where parameter loading is fully overlapped with computation, maximizing efficiency.
As shown, the optimal remapping percentage varies with changes in the runtime batch size, reflecting shifts in system load.
Using a static remapping percentage may result in suboptimal resource use: underutilizing GPU memory at off-peak load and causing compute stalls at peak load. 
For example, a 30\% remapping percentage tuned for average batch size in ShareGPT results in CPU-GPU bandwidth becoming a bottleneck during off-peak workloads.
Therefore, the remapping percentage must be dynamically adapted at runtime.

\noindent\textbf{Which layers to target?}
Once the remapping percentage is set, the next design challenge is selecting specific layers to remap.
This remains non-trivial, as \LLMPipe must decide how to allocate limited GPU memory across all layers -- deciding which layers remain permanently in GPU memory and which are transferred on the fly.
For an efficient pipeline, a layer selection strategy is needed to guide the layer-sharing mechanism in \LLMPipe.
A suboptimal layer selection can lead to reduced serving throughput (by 12.7\%, \S~\ref{section_effective_layer}). \par

%% file: 04_design.tex
We introduce \LLMPipe, a \textit{Dynamic Remapping Engine} for multi-tenant LLM inference serving systems that elastically adjusts the KV cache size by remapping parameter memory as KV cache memory. 
Figure~\ref{fig_overview} shows the \LLMPipe architecture, and depicts how \LLMPipe can be integrated into existing LLM inference serving systems. \par

\vspace{-2mm}
\subsection{Architecture of \LLMPipe}
\begin{figure}[t]
    \centerline{\includegraphics[width=0.9\columnwidth, trim = 39mm 41mm 79mm 44mm, page=6, clip=true]{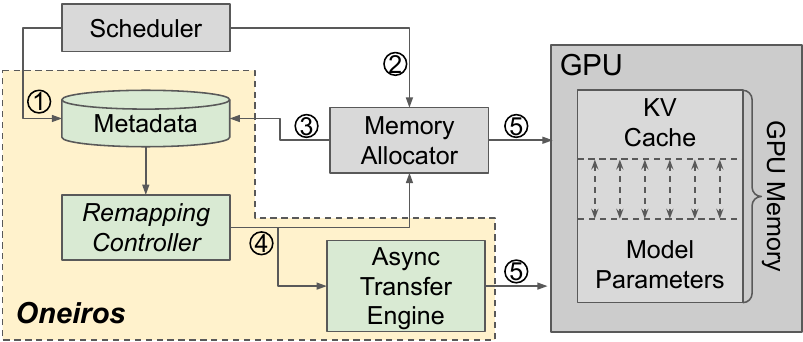}}
    \vspace{-4mm}
	\caption{Overview of \LLMPipe. }
    \vspace{-4mm}
	\label{fig_overview}
\end{figure}

\LLMPipe is designed to have three key components, marked in green color in Figure~\ref{fig_overview}.

\noindent{\textbf{Metadata Store:}}
The \textit{Metadata Store} maintains the information about the models and their memory utilization.
Model information includes the currently scheduled models (referred to as active models) and idle models currently not receiving requests (referred to as inactive models). 
Memory utilization information captures KV cache usage, with a focus on the amount of available free memory.

\noindent{\textbf{Remapping Controller:}}
The \Engine dynamically repurposes GPU memory allocated to inactive model parameters for use as KV cache by active models. 
When KV cache demand exceeds available GPU memory, the \Engine reclaims a portion of the parameter memory to expand KV cache capacity. 
Conversely, when KV cache usage subsides, the reclaimed memory is remapped back for parameter storage. \par

\noindent{\textbf{Async Transfer Engine:}}
The \textit{Async Transfer Engine} manages parameter loading and synchronizes CPU-GPU data transfers with GPU computation.
When the \Engine initiates parameter remapping, a portion of GPU memory originally allocated for model parameters is repurposed for KV cache usage. 
If the reclaimed parameters are later required for computation, the \textit{Async Transfer Engine} fetches them back from CPU memory to GPU memory.
By leveraging the deterministic execution order of layers within each \TokenGen iteration, the \textit{Async Transfer Engine} overlaps CPU-GPU data transfers with ongoing GPU computation. 
\par

\vspace{-1mm}
\subsection{Workflow}
\vspace{-1mm}
Figure~\ref{fig_overview} depicts the workflow of the interaction between various components to achieve the goal of dynamic parameter remapping in \LLMPipe.
The \textit{Scheduler} selects which models to serve, referred to as active models. 
Once active models are chosen, the \textit{Memory Allocator} updates the parameter and KV cache memory usage accordingly. 
Using both model information and KV cache utilization data, the \Engine determines whether parameter memory needs to be remapped to expand KV cache capacity. 
If remapping is required, the \Engine updates memory usage records in the \textit{Memory Allocator} and triggers the \textit{Async Transfer Engine} to initiate data transfer. 
The GPU then executes the LLM inference tasks.
With support from \LLMPipe, a portion of parameter memory in GPUs can be dynamically reclaimed at runtime, allowing the system to accommodate larger batches or longer sequences.

%% file: 05_00_engine.tex
\begin{figure*}[t]
    \centerline{\includegraphics[width=2.1\columnwidth, trim = 140mm 43mm 52mm 48mm, page=1, clip=true]{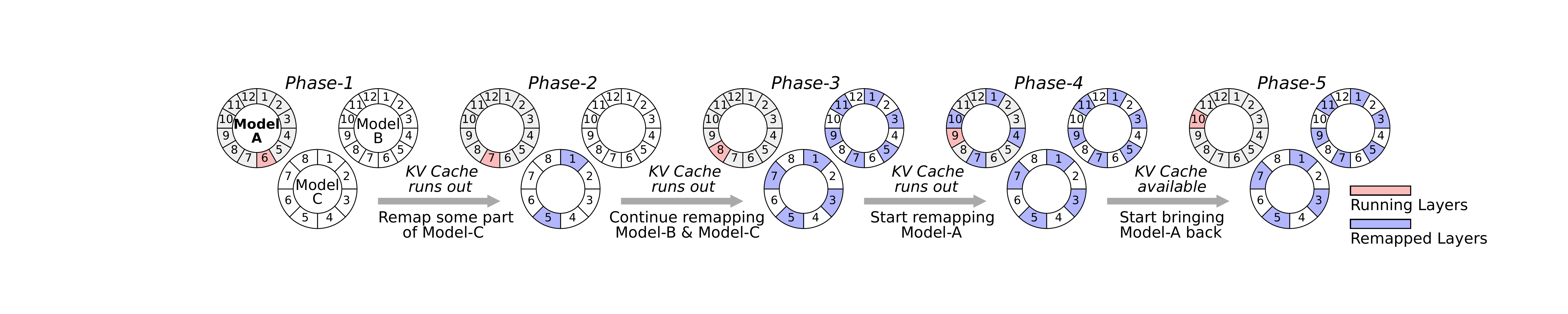}}
    \vspace{-4mm}
	\caption{\Engine decides which model and layer to remap based on runtime KV cache usage (Model-A is active). }
	\label{fig_remapping_layer}
    \vspace{-2mm}
\end{figure*}

\LLMPipe enables dynamic, runtime management of GPU memory through its \Engine. 
Figure~\ref{fig_remapping_layer} illustrates an example execution workflow of the \Engine across five phases, each representing a snapshot of model activity within an LLM serving system.
Depending on runtime conditions, the \Engine decides either to initiate remapping, continue remapping additional layers from the active or inactive models, or halt remapping altogether (reclaimed KV cache memory restored for parameter storage).
In this section, we describe how \Engine answers the key questions, including when to remap (\S~\ref{section_design_remapping_when}), which models to remap (\S~\ref{section_design_remapping_which_model}), how many (\S~\ref{section_design_remapping_how}), and which layers to remap (\S~\ref{section_design_remapping_which_layer}). 
Putting it all together, we then present an algorithm that can seamlessly integrate \LLMPipe into existing LLM inference serving frameworks, enabling dynamic remapping of model parameter memory to automatically and elastically expand KV cache capacity at runtime (\S~\ref{section_all_together}). \par 

\input{05_01}
\input{05_02}
\input{05_03}
\input{05_04}
\input{05_05}

%% file: 05_01.tex
\vspace{-3mm}
\subsection{When to Remap?}
\vspace{-1mm}
\label{section_design_remapping_when}
The \Engine initiates parameter remapping when the KV cache capacity is exhausted. 
While unidirectional parameter remapping incurs lower overhead than bidirectional KV cache swapping, \LLMPipe activates it only when necessary. 
As such, accurately determining both the initiation and termination points of remapping is critical. 
Given the dynamic nature of real-world production workloads~\cite{cortez2017resource, hadary2020protean, shahrad2020serverless, stojkovic2024smartoclock, patel2024splitwise, stojkovic2024dynamollm}, the \Engine enables remapping during peak request periods to accommodate larger batch sizes. 
During non-peak periods, when KV cache space is sufficient, remapping is disabled to avoid the overhead of reloading previously evicted parameters into GPU memory.
The reclaimed KV cache memory is then restored for parameter usage. \par

%% file: 05_02.tex
\vspace{-3mm}
\subsection{Which Models to Remap?}
\label{section_design_remapping_which_model}
\vspace{-1mm}
This section explains how the \Engine determines which models to remap.
We begin by reviewing existing GPU sharing techniques used in multi-tenant inference serving systems, as the \Engine adapts its model selection strategy based on the specific sharing mechanism employed.
We then present a detailed explanation of our model selection strategy. \par

\noindent{\textbf{GPU sharing background:}}
Given the memory capacity of emerging GPUs, enabling GPU sharing is essential to fully utilize their processing capabilities.
Prior works have examined both temporal~\cite{bai2020pipeswitch, gujarati2020serving, xiao2020antman, fu2024serverlessllm, patke2024queue} (example in Figure~\ref{fig_without_llm_pipe_1}) and spatial~\cite{dhakal2020gslice, romero2021infaas, lee2024parvagpu, strati2024orion, choi2022serving, xu2022igniter, han2024kace, wang2024improving} (example in Figure~\ref{fig_without_llm_pipe_2}) sharing strategies for efficient machine learning inference serving on GPUs.
Temporal sharing allocates the entire GPU to one model at a time, allowing for larger batches per model. 
This approach is especially useful in multi-agent workflows -- where the output of one model becomes the input to another~\cite{talebirad2023multi, chan2023chateval, li2024more, han2024llm, kim2025cost} -- and in cloud production environments, where models often experience idle periods with no incoming requests~\cite{yu2025prism}.
In contrast, spatial sharing divides GPU resources among multiple models simultaneously, allowing for faster response times at the expense of smaller batch sizes per model~\cite{lee2024parvagpu, duan2024muxserve}.
\textit{Each GPU resource sharing strategy is tailored to specific use cases and can be selectively employed by multi-tenant LLM inference schedulers.
The \Engine in \LLMPipe is designed to be compatible with any scheduling policy, although the design of the scheduler itself is beyond the scope of this study.
}

\noindent{\textbf{Model selection strategy:}}
We now present the model selection strategy to determine which models to remap parameter layers from. 
The \Engine prioritizes remapping parameters from inactive models with the lowest priority, as determined by the scheduling policy, rather than from active ones.
Under temporal GPU resource sharing, models scheduled for future execution are considered inactive.
In contrast, under spatial sharing, all models are treated as active, requiring each to remap parameters.
The number of layers to remap and their selection are determined using the strategies described in \S~\ref{section_design_remapping_how} and \S~\ref{section_design_remapping_which_layer}.
If the scheduler does not specify model priorities or all models are active, the \Engine defaults to a round-robin policy, treating all models equally, thereby ensuring fairness across models.
In this case, it applies a Most Recently Used (MRU) strategy, remapping parameters from the most recently activated model to defer parameter transfer costs as far into the future as possible. To avoid any starvation, \LLMPipe also caps the maximum number of layers remapped based on a threshold derived empirically through runtime profiling (more in \S~\ref{section_design_remapping_how}), and no model is ever remapped in its entirety.
This strategy is especially effective when future request patterns are unpredictable or when model priorities are unspecified. \par

\noindent{\textbf{Example:}}
Phase 2 of Figure~\ref{fig_remapping_layer} illustrates an example of the model selection policy.
In this scenario, Model-A is active, while Model-B and Model-C are inactive, with the scheduler assigning the lowest priority to Model-C.
The \Engine first evicts and remaps parameters from Model-C. 
To avoid starving inactive models, a maximum remapping threshold is enforced (see \S~\ref{section_design_remapping_how}), ensuring that enough parameters remain in GPU memory to support cold-start loading during the \Prompt phase when a model becomes active.
After reaching the remapping limit for Model-C, the \Engine continues with Model-B. Only once all inactive models have reached their remapping thresholds does the \Engine begin remapping parameters from the active Model-A. \par

%% file: 05_03.tex
\vspace{-3mm}
\subsection{How Many Layers to Remap?}
\label{section_design_remapping_how}
To hide parameter transfer latency by overlapping it with GPU computation, the transfer from CPU to GPU must finish before the computation completes; otherwise, it becomes a performance bottleneck.
Since both model parameter sizes and CPU-GPU bandwidth are known ahead of runtime, the transfer time per layer can be profiled offline and treated as prior knowledge by the online inference system. 
As a result, the \Engine only needs to monitor GPU computation time, denoted as $T_{Compute}$, at runtime to determine whether additional parameters can be remapped. 
For an active model, $T_{Compute}$ refers to the time required to generate the current token during the \TokenGen phase. 
For an inactive model, it represents the duration of the \Prompt phase, which is more compute-intensive and typically longer, as no tokens have been generated yet~\cite{patel2024splitwise, agrawal2024taming}.
Let $T_T$ denote the time required to transfer a single layer of parameters. 
To prevent parameter loading from becoming a new bottleneck, the number of remapped layers $N$ must satisfy the constraint $T_T \cdot N \leq T_{Compute}$. \par

The longer $T_{Compute}$ in the \Prompt phase allows more layers in inactive models to be remapped compared to when the model is active. 
Similar to the decode phase, the \Prompt phase executes one layer at a time rather than all at once. 
If all the parameters are not present in GPU memory because of the parameter remapping triggered by \LLMPipe, the \Engine transfers the parameters for the later layers from CPU memory back into the GPU memory while initial layers are being executed. 
Thus, the parameters are ready in GPU memory when needed, as the \Prompt phase makes progress layer-by-layer. 
The \Prompt phase is compute-bound, and takes additional compute time that provides an opportunity for \LLMPipe to hide the latency of loading remapped parameters back to GPU memory.
\par

%% file: 05_04.tex
\vspace{-3mm}
\subsection{Which Layers to Target?}
\label{section_design_remapping_which_layer}
This section details the layer selection strategy in the \Engine.
We begin by describing the strategy itself, followed by an example to illustrate its behavior.
Finally, we provide a theoretical proof to demonstrate that the strategy we used is optimal. \par

\noindent{\textbf{Layer selection strategy:}}
Understanding the layer selection strategy requires recognizing the circular execution pattern of layers in LLM inference, as shown in Figure~\ref{fig_remapping_layer}. 
The auto-regressive nature of LLM inference makes each of the subsequent tokens execute the same layers again. 
Thus, LLM inference can be viewed as a circular execution of layers, where the execution of the last layer for the current token is succeeded by the execution of the first layer for the next token.
The \Engine uses a uniform-interval layer selection strategy, in which evenly spaced layers are chosen to share a common GPU memory region allocated for their parameters.
These layers alternate access to the shared region -- at any given moment, some layer parameters reside in GPU memory, while others remain in CPU memory.
When a layer not currently in GPU memory is needed, its parameters are loaded into the shared region, replacing those of the previously loaded layer. \par

\noindent{\textbf{Example:}}
To provide a more intuitive explanation, we use the example of Model-C with 8 layers (Phase 2 of Figure~\ref{fig_remapping_layer}).  
The \Engine employs a uniform-interval layer selection strategy. 
For instance, when remapping the parameters of a single layer, Layers 1 and 5 are selected to share a common GPU memory region (the other 6 layers remain in GPU memory permanently), alternating between CPU and GPU memory. 
The uniform-interval layer selection strategy is designed to balance serving latency across successive token generations. 
In contrast, ignoring the circular nature of LLM inference -- by treating each token generation as an independent execution -- may lead to suboptimal choices, such as selecting Layer 1 and Layer 8. 
Although this configuration allows sufficient time to load the parameters of Layer 8 after executing Layer 1, it does not provide enough time to reload the parameters of Layer 1 before it is needed again. 
This demonstrates the benefit of uniform-interval selection in circular execution, which ensures that each remapped layer can be restored to GPU memory in time for reuse. \par

\noindent{\textbf{Theoretical proof:}}
Consider an \textit{n}-layer model in which the \Engine remaps the parameter memory of $\alpha$ layers for KV cache.
Consequently, all $n$ layers must be scheduled within the GPU memory footprint of only $n-\alpha$ layers. 
Suppose $m$ layers are transferred between CPU and GPU during each token generation. 
We note $L_1, L_2, \ldots, L_{m}$ as the $m$ layers (remapped layers in Figure~\ref{fig_remapping_layer}) that are transferred from CPU to GPU. 
We define $T_T$ as the time taken to load parameters of a single layer from CPU to GPU memory, and $T_{c}$ as the time spent on computation for a single layer on GPU.
\textit{Our layer selection policy aims to satisfy two key objectives: (1) ensure that parameter transfer latency does not become a performance bottleneck; and  (2) minimize CPU-GPU data transfer traffic. }
\par

\textit{To satisfy objective (1)}:
the CPU-to-GPU parameter loading must be fully overlapped with GPU computation.
During each token generation, the \Engine transfers $m$ layers from CPU to GPU memory. 
The goal is to select $m$ layers such that the computation between their executions is sufficient to hide the associated transfer latency.
For example, after layer $L_1$ completes execution, the \Engine begins transferring parameters of layer $L_2$. 
The cumulative computation time of the layers executed between $L_1$ and $L_2$ must be long enough to mask the transfer time of $L_2$.
Similarly, each subsequent transfer must also meet the condition to hide its transfer time, continuing until the parameters for layer $L_1$ (for the next token) are fully transferred.
We formally present this constraint as:
\begin{equation}
\label{eq:proof-1}
T_T \le \sum_{i=L_{1}}^{L_{2}} T_c\; \cdots, \; T_T \le \sum_{i=L_{m-1}}^{L_{m}} T_c, \ T_T \le \sum_{i=L_{m}}^{L_1} T_c
\end{equation}

\noindent Consider that there are $k_1$ layers between $L_1$ and $L_2$\footnote{In this paper, we assume all layers are identical. For emerging models with heterogeneity across layers~\cite{zhang2025jenga}, these formulas can be easily adapted, e.g., $T_T$ can be written as $T_{T,i}$, where $i$ denotes each specific layer, and solved using integer linear programming.}, and $k_{m}$ layers between $L_{m}$ and $L_{1}$ (for the next token), we have:
\begin{equation}
\label{eq:proof-2}
\sum_{i=L_1}^{L_2} T_c = k_1 \times T_c,
\; \cdots \;,
\sum_{i=L_{m}}^{L_{1}} T_c = k_{m} \times T_c
\end{equation}

\noindent Leveraging Equations \ref{eq:proof-1} and \ref{eq:proof-2}, we conclude that:
\begin{equation}
\label{eq:proof-3}
T_T  \le \min (k_1 \times T_c, \;\cdots,\; k_{m-1} \times T_c, \; k_{m} \times T_c)
\end{equation}

\noindent The Right-Hand Side (RHS) of Equation \ref{eq:proof-3} denotes the threshold beyond which transfer time becomes a performance bottleneck. 
To maximize the RHS under the constraint $k_1+k_2+\;\cdots\;+k_{m}=n$, the optimal solution is achieved when $k_1 = k_2 = \cdots = k_{m}$.
In other words, \textit{within the circular layer execution model, the selected $m$ layers must be evenly spaced, ensuring uniform intervals between consecutive remapped layers.} 
This aligns with the geometric principle that, for $n$ points on a circle, the configuration maximizing the minimum pairwise distance is achieved by placing them at equal angular intervals. \par

\textit{To satisfy objective (2)}: 
we aim to find the smallest valid integer value of $m$.
We begin with $m=\alpha$, meaning that only $\alpha$ layers are loaded from CPU memory during each token generation. 
However, this configuration is infeasible because loading a remapped layer into GPU memory requires evicting a resident layer, which will be needed in the next token generation due to the circular reuse pattern of all layers.
As a result, $m=\alpha$ is insufficient, and the minimum viable value becomes $m=\alpha+\beta$ where $\beta \ge 1$. 
In this configuration, $n-(\alpha+\beta)$ layers remain permanently in the GPU memory, while the remaining $\alpha+\beta$ layers share a GPU memory region sized for $\beta$ layers. \par

We start with $\beta = 1$. 
To satisfy design objective (1), we have Equation~\ref{eq:proof-3} to make the data transfer time shorter than the computation time.
However, the prerequisite of Equation~\ref{eq:proof-3} is that parameter transfers from CPU to GPU can be initiated at any time, overlooking the data dependencies between layers: the $\alpha+1$ layers share a single GPU memory slot, meaning that the transfer of a layer cannot begin until its computation has completed. Accounting for this constraint, the time budget becomes more restrictive:
\begin{equation}
\label{eq:proof-4}
T_T \times (\alpha + 1) \le T_c \times (n -\alpha - 1) 
\end{equation}
To relax this constraint, one possible solution is to adopt a double-buffering approach, increasing the number of layers transferred on the fly from $\alpha + 1$ to $\alpha + 2$.
With two GPU memory slots, these $\alpha + 2$ layers can be transferred and computed in a staggered manner, allowing the transfer of the next layer to begin without waiting for the execution of the current layer to complete.
This results in a less restrictive timing condition:
\begin{equation}
\label{eq:proof-5}
T_T \times (\alpha + 2) \le T_c \times n 
\end{equation}

When $\alpha$ is small and $n$ is large, setting $m = \alpha + 1$ is generally preferred as it minimizes CPU-GPU transfer. 
In contrast, $m = \alpha + 2$ is more suitable for larger models with more remapped parameters.
For instance, based on Equations~\ref{eq:proof-4} and ~\ref{eq:proof-5}, when $\alpha \ge 9$ for a model with 40 layers ($n=40$), choosing $m = \alpha + 2$ can provide better performance (\S~\ref{section_effective_layer}). \par

%% file: 05_05.tex
\vspace{-2mm}
\subsection{Putting All Together}
\label{section_all_together}
Algorithm~\ref{algo_remapping} outlines the complete execution workflow of the \Engine. 
Consistent with existing LLM inference serving schedulers, the \Engine makes parameter remapping decisions at a per-token granularity. 
At each iteration step, it first evaluates the runtime status of the \textit{Memory Allocator}, to determine whether remapping needs to be triggered (Line 3).
Then, it marks the GPU memory previously occupied by parameters as available for KV cache. 
The \Engine then updates the list of model layers whose parameters will be remapped between CPU and GPU memory (Line 4). 
As illustrated in the remapping() function of Algorithm~\ref{algo_remapping}, the \Engine continuously tracks the current remapping percentage for each model (Lines 18-23) and prioritizes remapping parameters from the model (Line 16) with the lowest priority decided by the scheduler. 
Meanwhile, when the KV cache space is freed after requests finish, the \Engine updates the remapping percentage based on the available memory information from the memory allocator (Lines 7-12). 
Finally, the \Engine updates parameter remapping information (remapped models and layers) to the GPU LLM Kernel, which executes the current iteration accordingly (Line 13). 
\par

\SetKwComment{Comment}{/* }{ */}
\SetKwFunction{FSchedule}{Remapping\_Controller}
\SetKwFunction{Fremapping}{Remapping}
\SetKwProg{Fn}{Function}{:}{}
\begin{algorithm}[t]
\DontPrintSemicolon
\footnotesize
\Fn{\FSchedule{}}{ 
    \ForEach{step() $\in$ LLM Inference Serving Scheduler} {
        \uIf{Running out of KV Cache blocks} {
            remapped\_layer\_list = remapping(); \\
            enable\_remap = True; \\
        }
        \uElse{
            \ForEach{Tensor $\in$ \textnormal{Added KV Cache}} {
                \uIf{Tensor is empty} {
                    remapped\_layer\_list = remapping(); \\
                }
            }
            \uIf{All Tensors are empty} {
                enable\_remap = False; \\
            }
        }
        GPU LLM Kernel(enable\_remap, remapped\_layer\_list); \\
    }
}

\Fn{\Fremapping{}}{ 
    model = pop(inactive\_model\_list); \\ 
    \uIf{model is not NULL} {
        model.remapped\_layers++; \\
        \uIf{model.remapped\_layers == model.layers} {
            inactive\_model\_list.remove(model); \\
        }
        sort(inactive\_model\_list); \\
    }
    \uElse{
        remapping parameters of active models;
    }
}

\caption{\Engine Workflow}
\label{algo_remapping}
\end{algorithm}

%% file: 06_implementation.tex
We integrate \LLMPipe into vLLM~\cite{kwon2023efficient}, a state-of-the-art LLM inference-serving framework. 
We add $\sim$$3,000$ lines of Python code and $\sim$$300$ lines of C++/CUDA code to enable support for parameter remapping engines and dynamic KV cache allocation.

\noindent{\textbf{Remapping Controller} and \textbf{Async Transfer Engine:}}
During inference, the \Engine coordinates with the \textit{Transfer Engine} to specify which layers from which models are to be remapped.
If remap is disabled, \LLMPipe follows the standard execution flow unchanged. 
When remap is enabled, the \textit{Transfer Engine} initiates an asynchronous transfer of the parameters of the target layer, overwriting\footnote{State-of-the-art LLM inference frameworks, such as vllm~\cite{kwon2023efficient}, keep a complete copy of all model parameters in CPU memory before transferring to GPU memory.} the GPU tensor memory previously used by layers that have completed execution.
To ensure correctness, the GPU performs memory barrier synchronization checks before launching kernels that depend on remapped parameters.

\noindent{\textbf{KV Cache Allocation:}}
Most state-of-the-art LLM frameworks, such as vLLM~\cite{kwon2023efficient}, support dynamic management of KV cache memory space. 
However, the physical memory for KV cache still needs to be statically allocated before serving starts. 
During parameter remapping, a portion of the parameter memory must be dynamically reconfigured as KV cache memory, requiring support for runtime dynamic KV cache allocation and deallocation.
To enable dynamic allocation and remapping of GPU memory, we adopt the vAttention design~\cite{prabhu2024vattention}, which reserves sufficient virtual memory space for the KV cache and maps physical memory pages on demand.
Once parameter tensors are released, the KV cache engine can immediately reuse the freed physical memory.
This approach also allows GPU physical KV cache memory to be efficiently shared across multiple models. \par

%% file: 07_00_evaluation.tex
We comprehensively evaluate the effectiveness of \LLMPipe by answering the questions below:

\begin{itemize}
    \item How effective is \LLMPipe in reducing tail latency and improving throughput under temporal GPU sharing? (\S~\ref{section_results_multi_online}) 
    \item Is \LLMPipe still effective in enhancing performance under spatial GPU resource sharing? (\S~\ref{section_results_single_online})
    \item How effective is the parameter remapping mechanism compared with KV cache swapping? (\S~\ref{section_results_vs_pie})
    \item How effective is the layer selection strategy used? (\S~\ref{section_effective_layer})
    \item How important is the decision of when to remap and how many layers to remap? (\S~\ref{section_ablation})
\end{itemize}

Some highlights of our results include:
\begin{itemize}
    \item \LLMPipe significantly improves multi-tenant LLM inference with temporal GPU sharing, reducing tail TBT latency by 44.8\%-82.5\%, tail TTFT latency by 74.8\%-99.3\%, and achieving 39.9\%-86.7\% higher throughput compared to vLLM.
    \item In addition to temporal sharing of GPU resources, \LLMPipe also supports spatial sharing. \LLMPipe achieves 20.7\%-65.5\% reduction in tail latency.
    \item Compared to KV cache swapping, \LLMPipe achieves 47.1\% higher throughput. 
    Additionally, it supports multi-tenant LLM inference serving, providing greater flexibility.
\end{itemize}

\input{07_01}
\input{07_02}
\input{07_03}
\input{07_04}
\input{07_05}
\input{07_06}

%% file: 07_01.tex
\begin{table}[t]
    \caption{Evaluated model combinations and their corresponding GPU memory reservations (\% of total GPU memory).}
    \label{tab:model_comb}
    \vspace{-4mm}
    \centering
    \setlength{\tabcolsep}{1pt}
    \renewcommand{\arraystretch}{1.25}
    \begin{tabular}{|| >{\centering\arraybackslash}p{1cm} |
                       >{\centering\arraybackslash}p{7cm} 
                       ||}
        \hline
        {\bf C1} & OPT-13b (35\%), Lllma-2-13b (35\%), Lllma-3-8b (20\%) \\
        \hline
        {\bf C2} & OPT-30b (65\%), OPT-6.7b (15\%) \\
        \hline
    \end{tabular}
\end{table}

\begin{figure*}[t]
  \centering
  \begin{subfigure}[b]{0.33\textwidth}
    \centering
    \includegraphics[width=0.9\textwidth, trim = 7mm 8mm 6mm 0mm, page=1, clip=true]{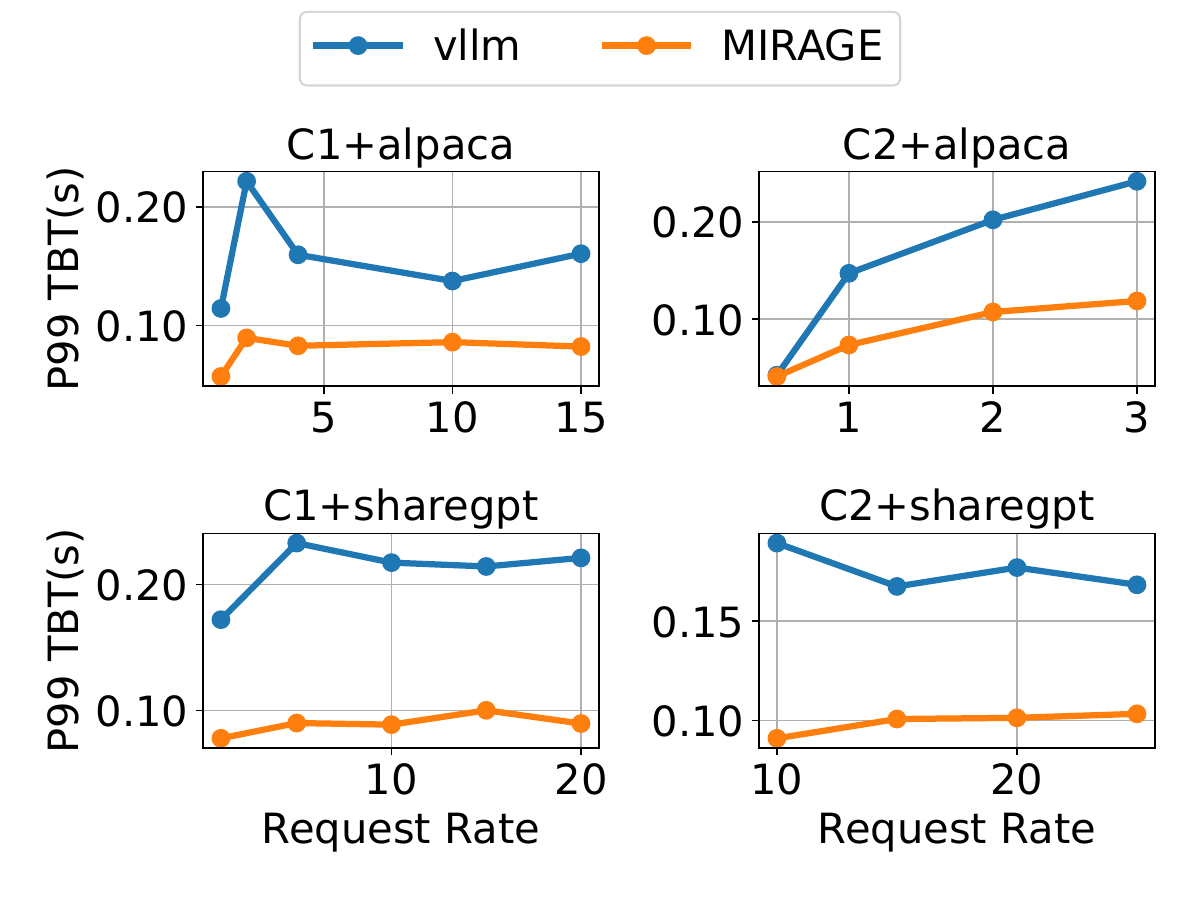}
    \vspace{-2mm}
    \caption{\LLMPipe reduces tail TBT latency.}
    \label{subfig:multi_tbt}
  \end{subfigure}
  \hfill
  \begin{subfigure}[b]{0.33\textwidth}
    \centering
    \includegraphics[width=0.9\textwidth, trim = 7mm 8mm 6mm 0mm, page=1, clip=true]{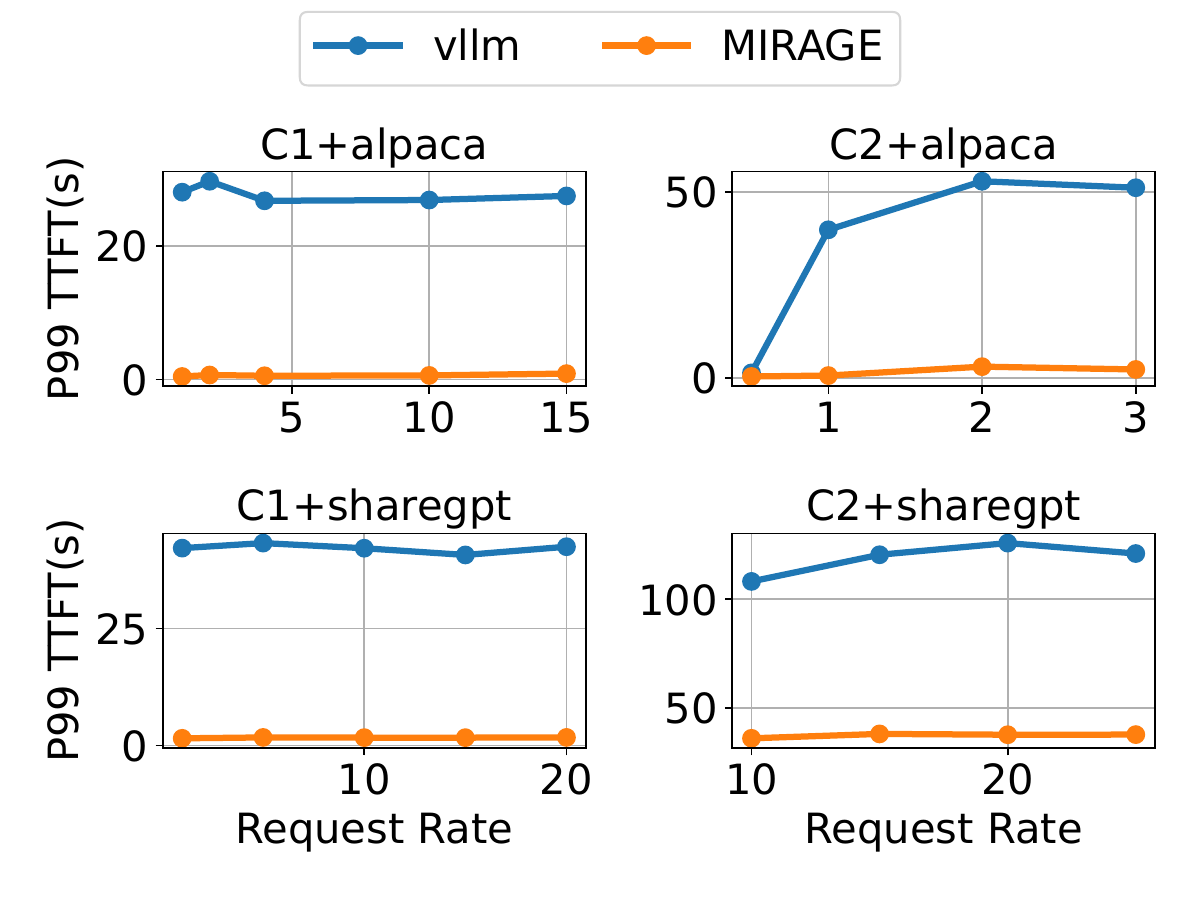}
    \vspace{-2mm}
    \caption{\LLMPipe reduces tail TTFT latency.}
    \label{subfig:multi_ttft}
  \end{subfigure}
  \hfill
  \begin{subfigure}[b]{0.33\textwidth}
    \centering
    \includegraphics[width=0.9\textwidth, trim = 7mm 8mm 6mm 0mm, page=1, clip=true]{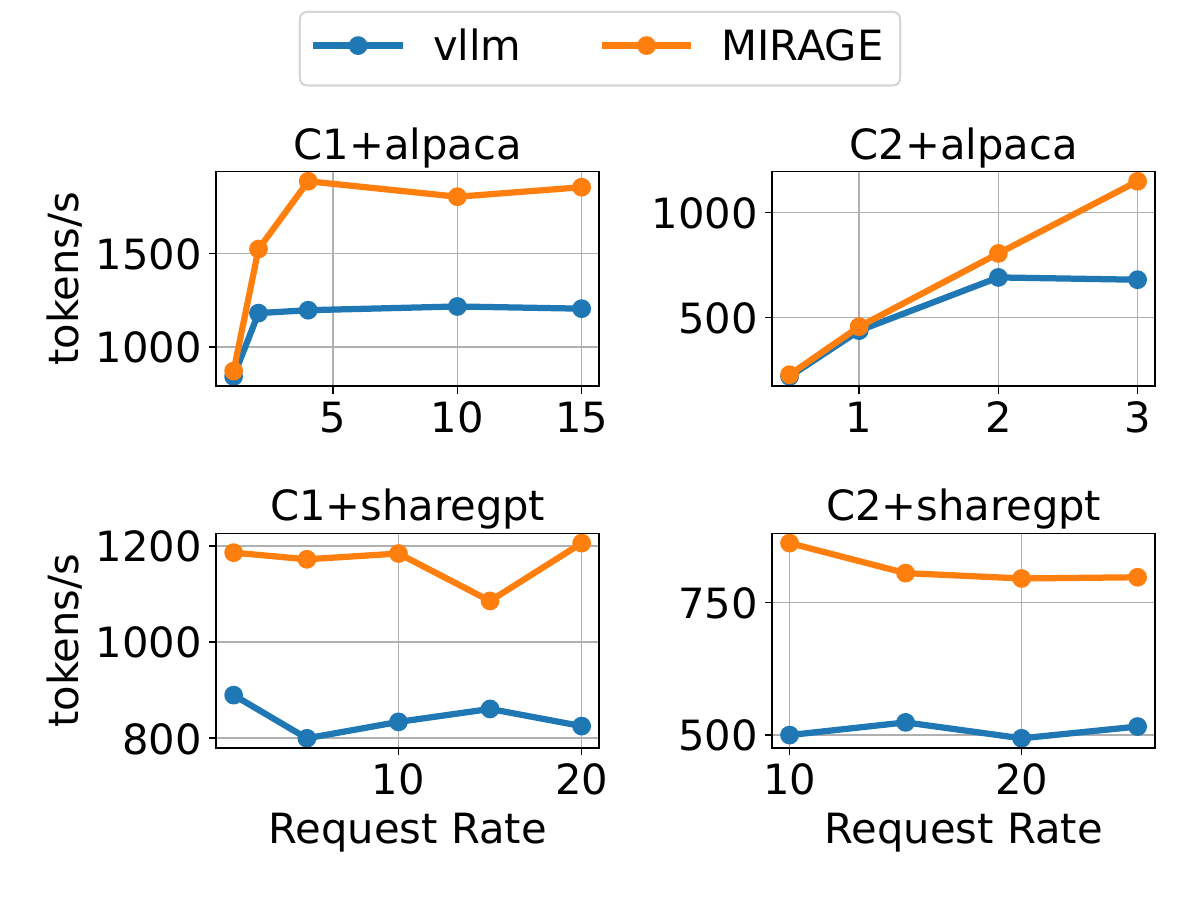}
    \vspace{-2mm}
    \caption{\LLMPipe improves throughput.}
    \label{subfig:multi_tp}
  \end{subfigure}
  \vspace{-8mm}
  \caption{\LLMPipe reduces latency and improves throughput compared with vllm on both C1 and C2.}
  \label{fig:multi_tenant_alp_sgpt}
\end{figure*}

\subsection{Experimental Setup}
\noindent{\textbf{Platforms:}}
We integrate \LLMPipe into vLLM v0.7.3~\cite{kwon2023efficient}, using CUDA 12.4, PyTorch 2.5.1, and Linux kernel 5.14. All experiments were conducted on the NVIDIA GH200 Grace Hopper Superchip, which features an H200 GPU with 96GB of HBM3 memory and a CPU host with 72 Arm Neoverse V2 cores and 224GB of LPDDR5X memory, connected via a 900 GB/s NVLink interconnect~\cite{GH200}. \par

\noindent{\textbf{Models:}}
We evaluated the following models: OPT~\cite{zhang2022opt} with 13b and 30b parameters; Llama-2~\cite{llama} with 13b parameters; and Llama-3~\cite{dubey2024llama} with 8b parameters. 
For multi-tenant serving, we evaluate two model combinations \textit{C1} and \textit{C2} in Table~\ref{tab:model_comb}.
\par

\noindent{\textbf{Traces:}}
We evaluate \LLMPipe using the Azure coding LLM inference traces that capture real-world query arrival patterns characterized by bursty query patterns~\cite{patel2024splitwise, stojkovic2024dynamollm}. 
Following prior work~\cite{ahmad2024proteus, gujarati2020serving, li2023alpaserve, miao2024spotserve}, we scale these traces to different query arrival rates while preserving their original characteristics, such as burstiness.

\noindent{\textbf{Datasets:}}
We use the ShareGPT~\cite{sharedgpt} and Alpaca~\cite{taori2023stanford} datasets, both of which contain input-output pairs from real LLM services. 
To simulate realistic multi-tenant scenarios, we additionally generate requests of varying lengths (synthetic datasets).

\noindent{\textbf{Baselines:}} 
We compare \LLMPipe against two baselines: vLLM~\cite{kwon2023efficient} and Pie~\cite{xu2024pie}.
vLLM is a state-of-the-art LLM inference serving system widely used in both academia and industry. 
Pie is the first to explore using the NVIDIA Grace Superchip for KV cache swapping. 

\noindent{\textbf{Metrics:}} 
Our evaluation focuses on two key metrics: (a) tail (P99) latency, which is critical for managing congestion in online systems~\cite{haque2015few, prekas2017zygos, delimitrou2018amdahl, kaffes2019shinjuku, zhao2023scalable}; we consider both TBT and TTFT latency; and (b) throughput (tokens per second), which quantifies the number of tokens a system can handle within a specific time frame.

%% file: 07_02.tex
\subsection{\LLMPipe Supports Temporal GPU Sharing}
\label{section_results_multi_online}
We first evaluate \LLMPipe under temporal GPU resource sharing, which is well-suited for multi-agent workflows and scenarios with frequent model idle periods. \par

\noindent{\textbf{Tail Latency:}}
We first evaluate \LLMPipe by comparing tail TBT and TTFT latencies.
Figures \ref{subfig:multi_tbt} and \ref{subfig:multi_ttft} present the P99 TBT and TTFT latency comparisons between \LLMPipe and vLLM. 
Across all scenarios, \LLMPipe consistently reduces tail latency. 
On average, when serving \textit{C1}, \LLMPipe achieves an average reduction of 54.4\% in P99 TBT latency and 96.7\% in P99 TTFT latency across both the Alpaca and ShareGPT datasets. 
Similarly, when serving \textit{C2}, \LLMPipe achieves an average reduction of 44.8\% in P99 TBT latency and 74.8\% in P99 TTFT latency on the same datasets.
\par

Without \LLMPipe, peak-load conditions that exhaust available KV cache typically cause existing serving systems~\cite{kwon2023efficient} to pause the \TokenGen process for all active requests, evict one, and recompute its KV cache to free space. 
\LLMPipe eliminates the need for recomputation, maintaining low tail latency, thereby ensuring stable performance for online inference serving. \par

\begin{figure}[t]
    \centerline{\includegraphics[width=0.9\columnwidth, trim = 9mm 8mm 8mm 10mm, page=1, clip=true]{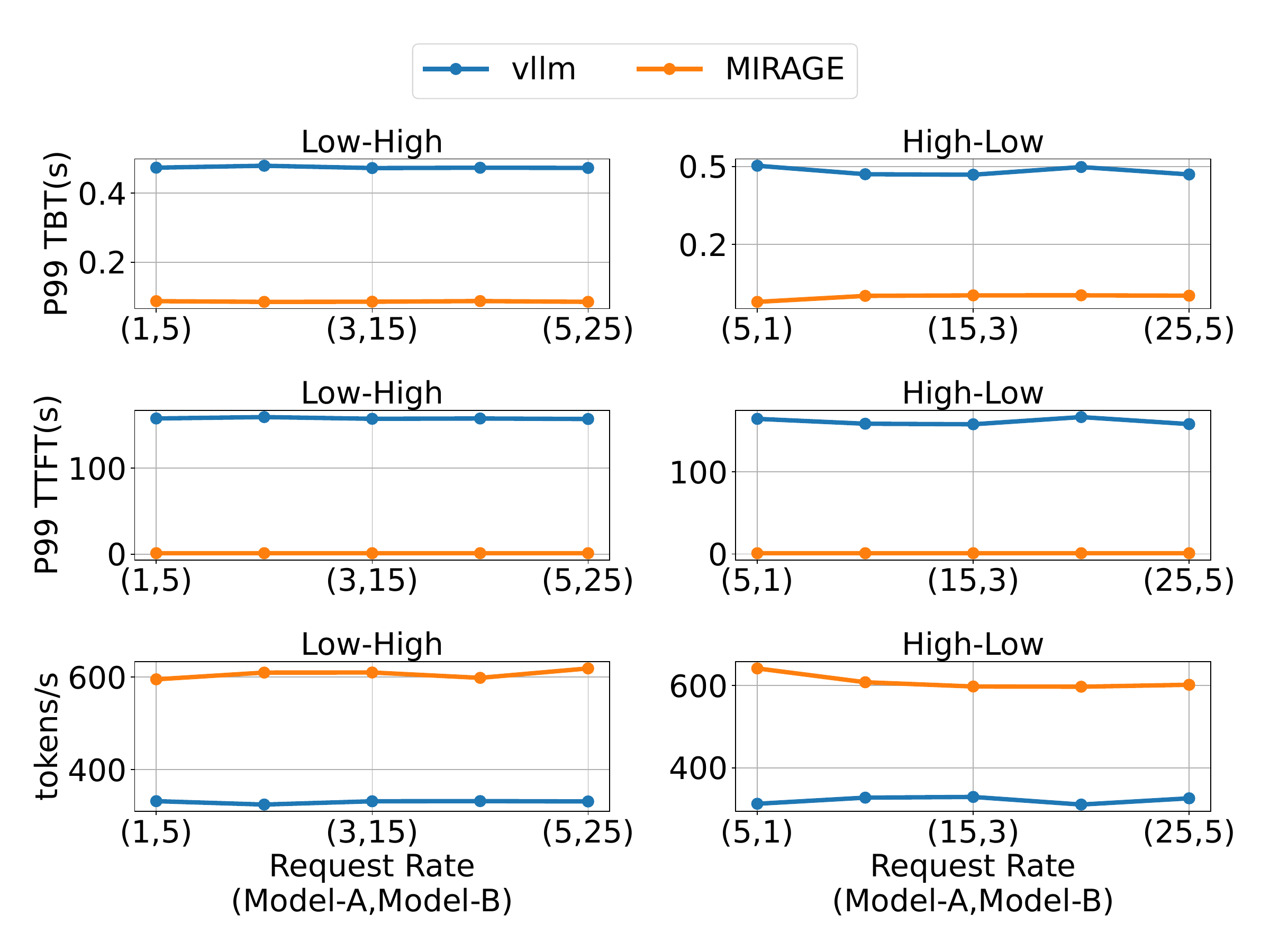}}
    \vspace{-4mm}
	\caption{
    \LLMPipe outperforms vLLM on OPT-30b (Model-A) + OPT-6.7b (Model-B) with different request arrival rates.
    }
	\label{fig:diff_rps}
\end{figure}

\noindent{\textbf{Throughput:}}
By using the \textit{Async Transfer Engine}, \LLMPipe overlaps parameter transfer with GPU computation. 
This not only reduces tail latency but also sustains high serving throughput.
Across all evaluated scenarios shown in Figure \ref{subfig:multi_tp}, \LLMPipe consistently increases throughput. 
On average, when serving \textit{C1}, it achieves a 39.9\% improvement in throughput across both the Alpaca and ShareGPT datasets. 
Similarly, when serving \textit{C2}, \LLMPipe delivers an average throughput gain of 45.3\% on the above datasets. \par

\begin{figure}[t]
    \centerline{\includegraphics[width=0.95\columnwidth, trim = 8mm 8mm 6mm 9mm, page=1, clip=true]{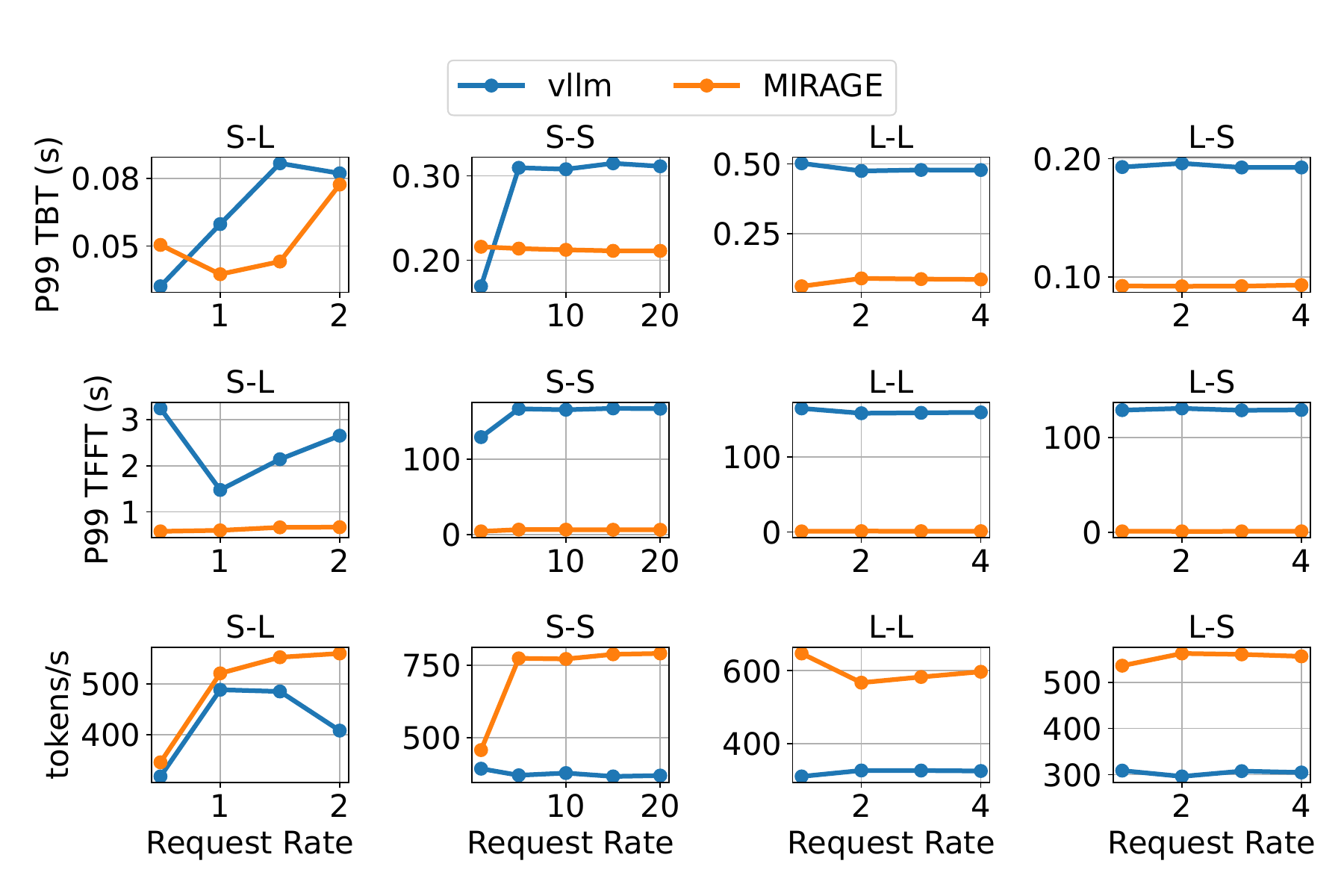}}
    \vspace{-4mm}
	\caption{\LLMPipe outperforms vLLM on C2 across both short (S) and long (L) input requests. }
	\label{fig:synth_ds}
\end{figure}

\noindent{\textbf{Varied Arrival Rates:}}
In real-world multi-tenant scenarios, different models or tasks often experience varying arrival rates~\cite{romero2021infaas, li2023alpaserve, sun2024llumnix, yu2025prism}.
\LLMPipe is designed to handle this variability effectively.
To demonstrate its effectiveness, we evaluate \LLMPipe by serving \textit{C2}, with different arrival rates across different models.
As shown in Figure~\ref{fig:diff_rps}, \LLMPipe consistently improves performance across all tested scenarios of varied input arrival rates, achieving an average of 86.7\% increase in throughput and 82.5\% and 99.3\% reduction in TBT and TTFT tail latencies, respectively. \par

\noindent{\textbf{Varied Inputs:}}
To emulate realistic multi-tenant scenarios -- where different models handle distinct tasks within a multi-agent flow and often operate on separate input datasets~\cite{li2024more, li2024survey, talebirad2023multi, sun2024llumnix} -- we construct arrival traces for two models (evaluated on \textit{C2}) using two request-type combinations: (a) synthetic long requests (average of 1734 tokens) paired with synthetic short requests (average of 634 tokens); and (b) synthetic short requests paired with synthetic long requests.
As shown in Figure \ref{fig:synth_ds}, \LLMPipe consistently improves performance across all evaluated scenarios, achieving an average throughput increase of 65.6\% and reductions of 57.1\% and 98.1\% in P99 TBT and TTFT latencies, respectively.
\par

\begin{figure}[t]
    \centerline{\includegraphics[width=\columnwidth, trim = 6mm 6mm 6mm 6mm, page=1, clip=true]{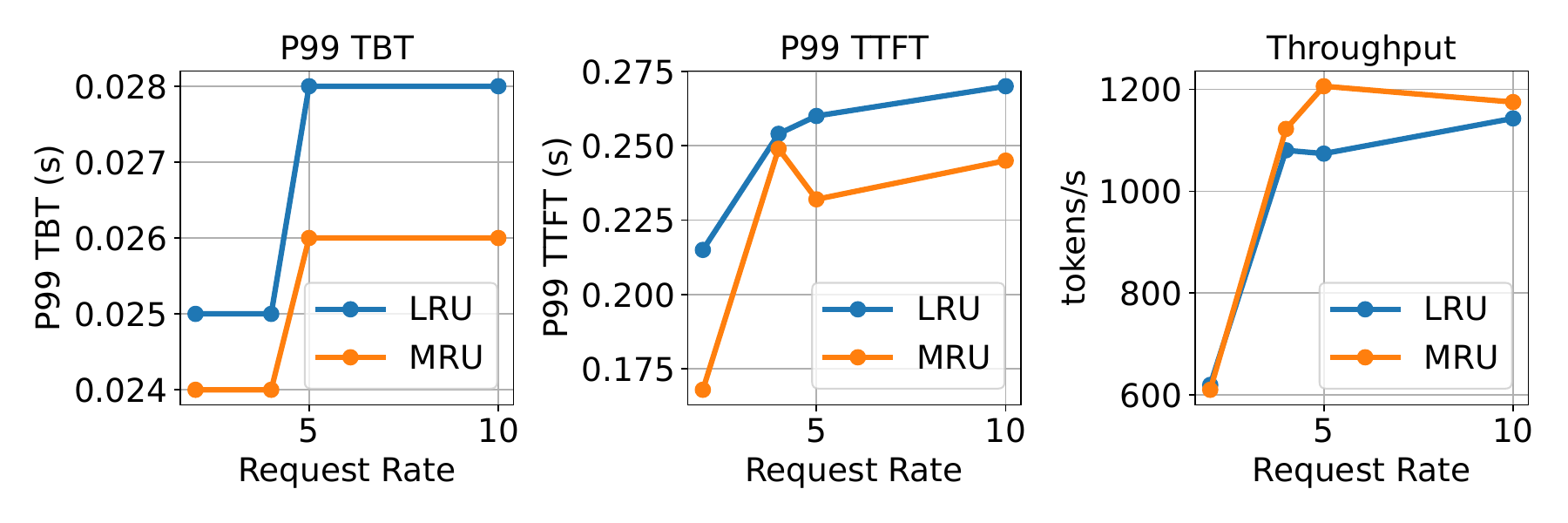}}
    \vspace{-4mm}
    \caption{MRU (used in \LLMPipe) remaps models expected to be used furthest in the future, reducing tail latency and improving throughput compared to LRU. }
    \label{fig:lru-vs-mru}
\end{figure}

\noindent{\textbf{Which models to remap?}}
We evaluate \Engine's model selection strategy (\S~\ref{section_design_remapping_which_model}) under temporal GPU sharing.
This strategy aligns with the multi-tenant scheduling policy that decides model execution order.
In the absence of explicit model priorities, \LLMPipe uses a default round-robin scheduling policy with a Most Recently Used (MRU) eviction strategy. 
To assess its effectiveness, we compare MRU against a Least Recently Used (LRU) policy, which remaps the parameters of the model most likely to be served next.
Figure~\ref{fig:lru-vs-mru} presents performance results for the \textit{C1} model combination under round-robin execution.
With the MRU policy, \LLMPipe achieves up to a 22.0\% reduction in tail latency compared to the LRU policy.
MRU consistently outperforms LRU by deferring transfer costs, remapping parameters of models that are less likely to be reused in the near future. \par

%% file: 07_03.tex
\vspace{-3mm}
\subsection{\LLMPipe Supports Spatial GPU Sharing}
\label{section_results_single_online}
In addition to supporting temporal GPU sharing, \LLMPipe also performs effectively under spatial sharing for online inference.
In this setup, all models remain active and maintain separate KV cache spaces.
To assess the effectiveness of \LLMPipe, we evaluate the model combination \textit{C1} on both Alpaca and ShareGPT datasets using Azure coding traces.
We evaluate the performance of \LLMPipe under two representative spatial sharing methods on NVIDIA-based GPU systems:
(1) \textit{Non-strict physical isolation}, as seen in Multi-Process Service (MPS)~\cite{MPS}, enables concurrent execution by allowing multiple processes to share internal GPU resources. 
It offers limited isolation since it does not partition components such as caches or memory controllers.
(2) \textit{Strict physical isolation}, exemplified by Multi-Instance GPU (MIG)~\cite{MIG}, partitions the GPU into independent instances, providing strong physical isolation across workloads.

\begin{figure}[t]
    \centerline{\includegraphics[width=\columnwidth, trim = 20mm 6mm 32mm 18mm, page=1, clip=true]{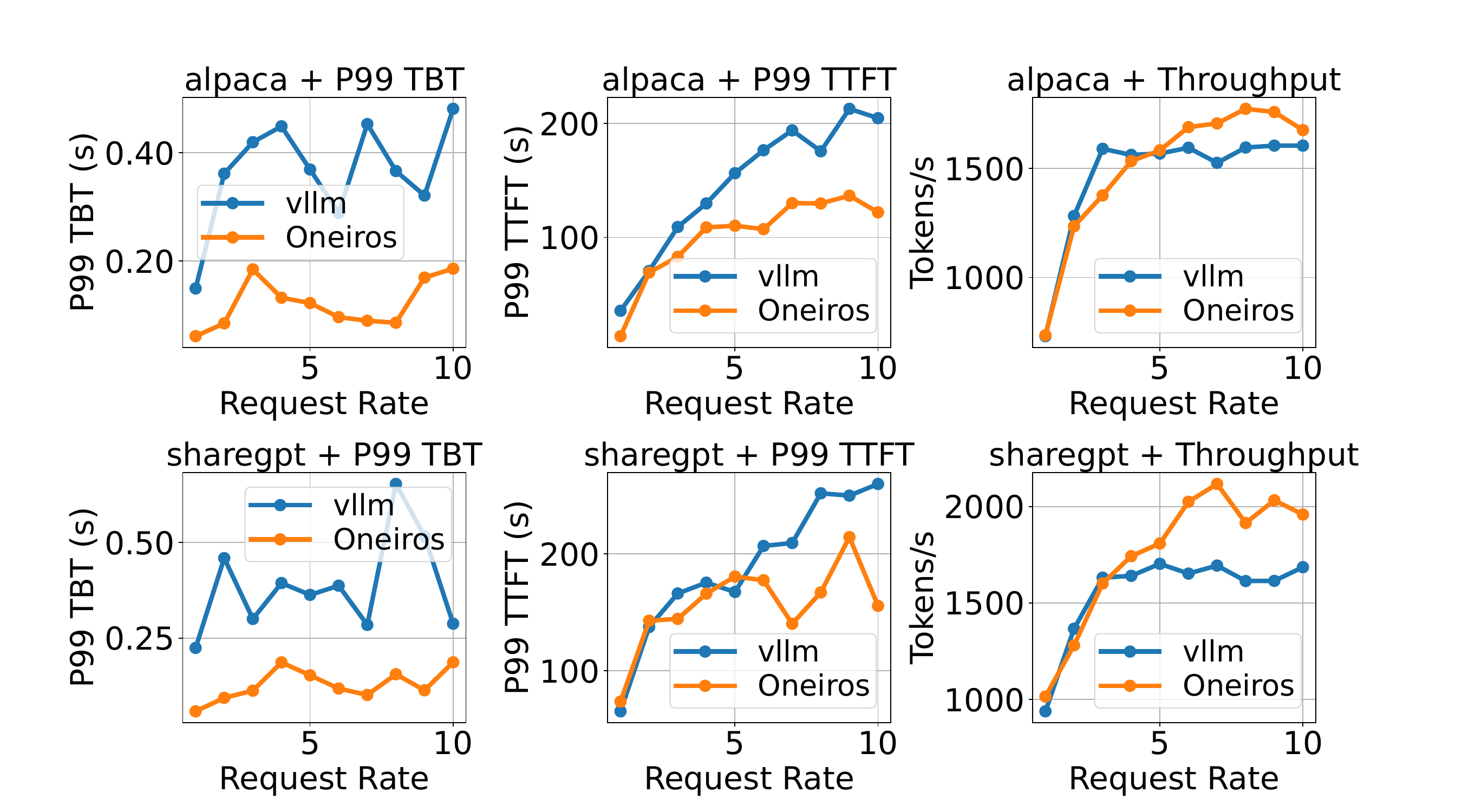}}
    \vspace{-4mm}
	\caption{\LLMPipe achieves higher throughput and lower tail TBT latency than vLLM under non-strict physical isolation. }
	\label{fig_spatial_mps}
\end{figure}

\noindent{\textbf{Non-Strict Physical Isolation:}}
As shown in Figure~\ref{fig_spatial_mps}, \LLMPipe effectively reduces both tail TBT and TTFT latency as request arrival rates increase toward peak load.
On average, it achieves a 65.5\% reduction in P99 TBT, a 20.7\% reduction in P99 TTFT, and a 6.6\% increase in throughput across varying arrival rates.
These results highlight the robustness of \LLMPipe: it not only delivers strong performance under temporal GPU resource sharing but also offers significant improvements in spatial sharing configurations. \par

\begin{figure*}[t]
	\begin{minipage}[t]{0.33\textwidth}
		\centering
		\includegraphics[width=\textwidth,trim = 10mm 6mm 24mm 22mm, page=1, clip=true]{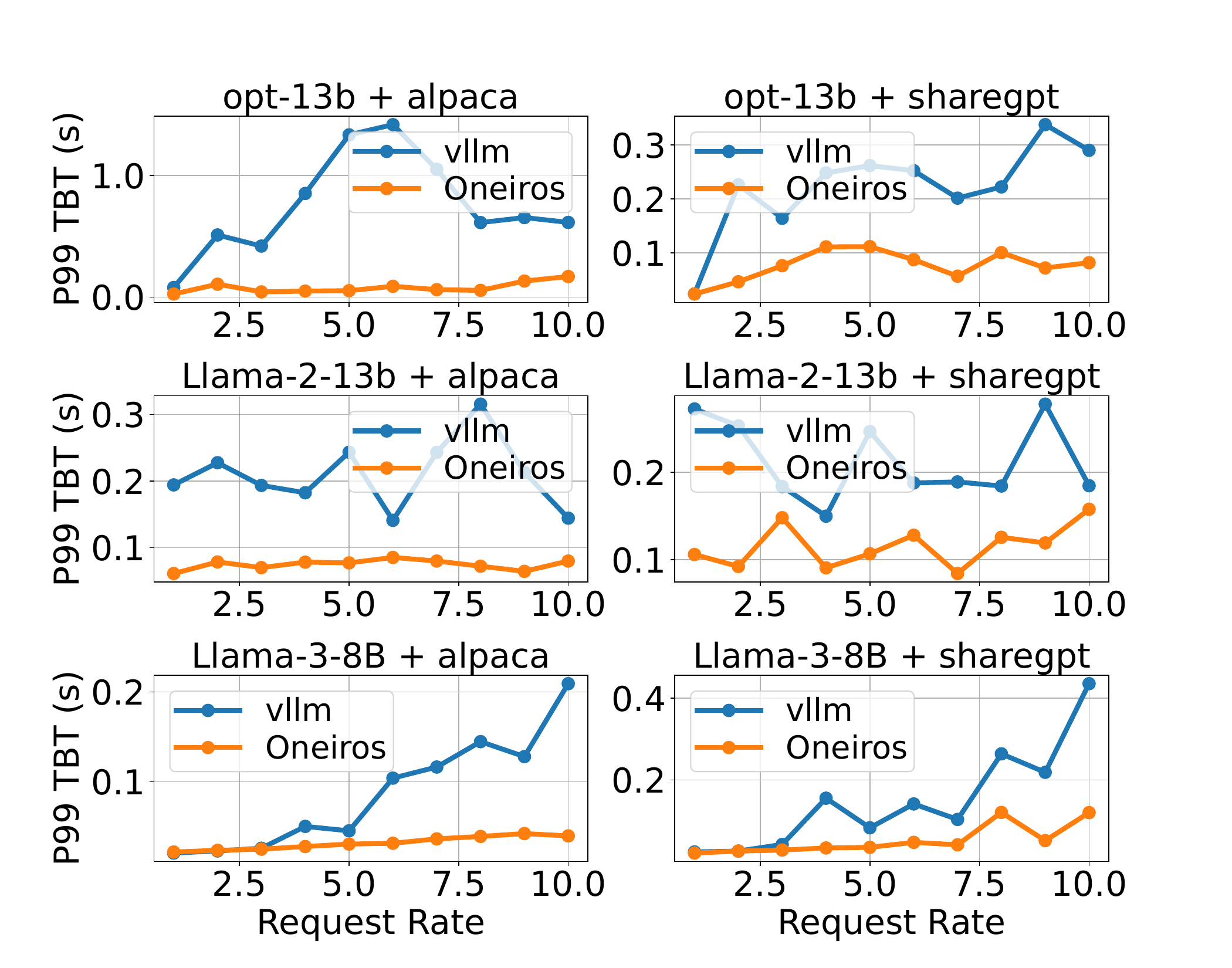}
        \vspace{-5mm}
		\subcaption{Tail TBT Latency.}
		\label{fig_spatial_mig_1}
	\end{minipage} 
	\begin{minipage}[t]{0.33\textwidth}
		\centering	\includegraphics[width=\textwidth,trim = 2mm 6mm 24mm 22mm, page=1, clip=true]{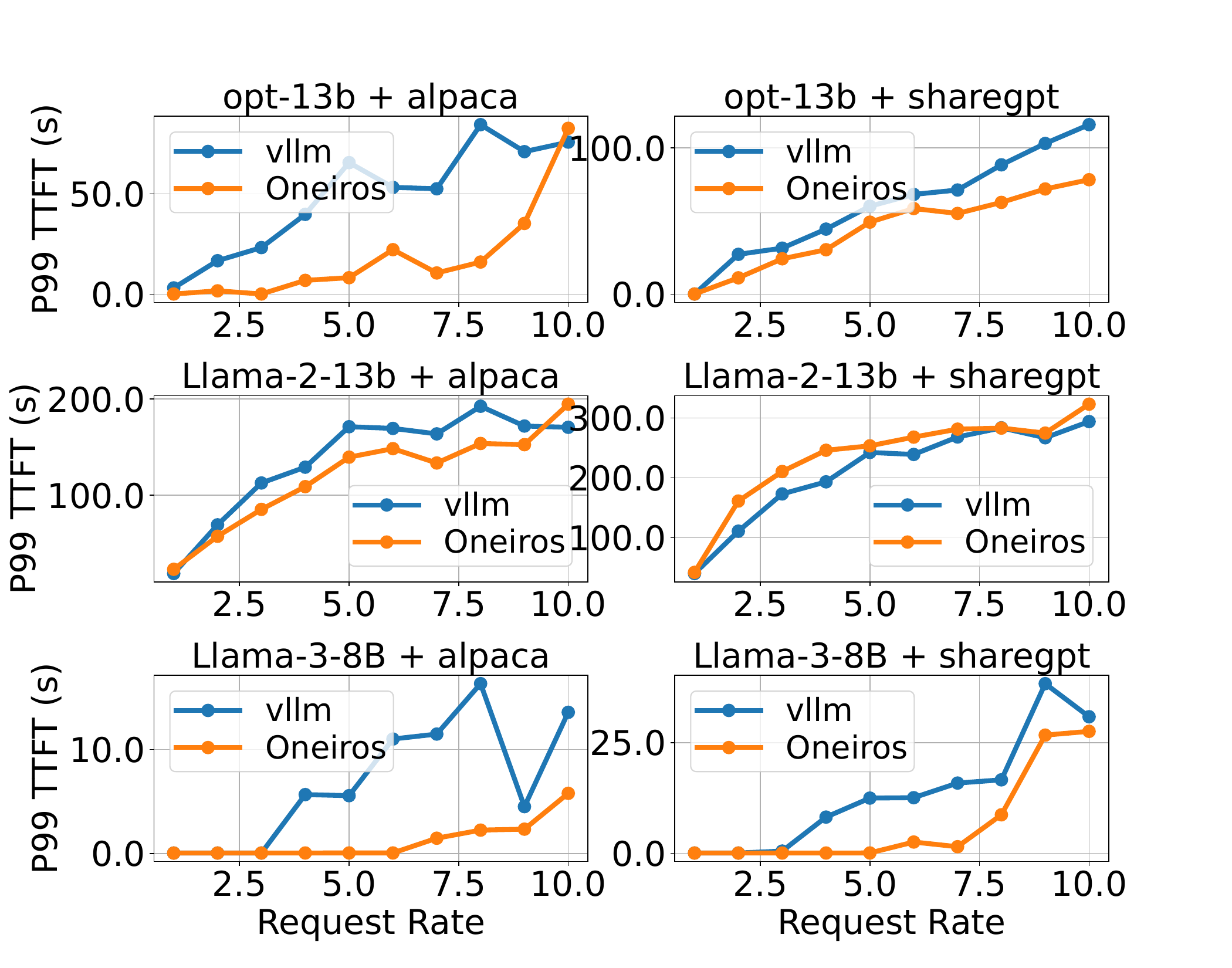}
        \vspace{-5mm}
		\subcaption{Tail TTFT Latency.}
		\label{fig_spatial_mig_2}
	\end{minipage}
    \begin{minipage}[t]{0.33\textwidth}
		\centering	\includegraphics[width=\textwidth,trim = 4mm 6mm 27mm 22mm, page=1, clip=true]{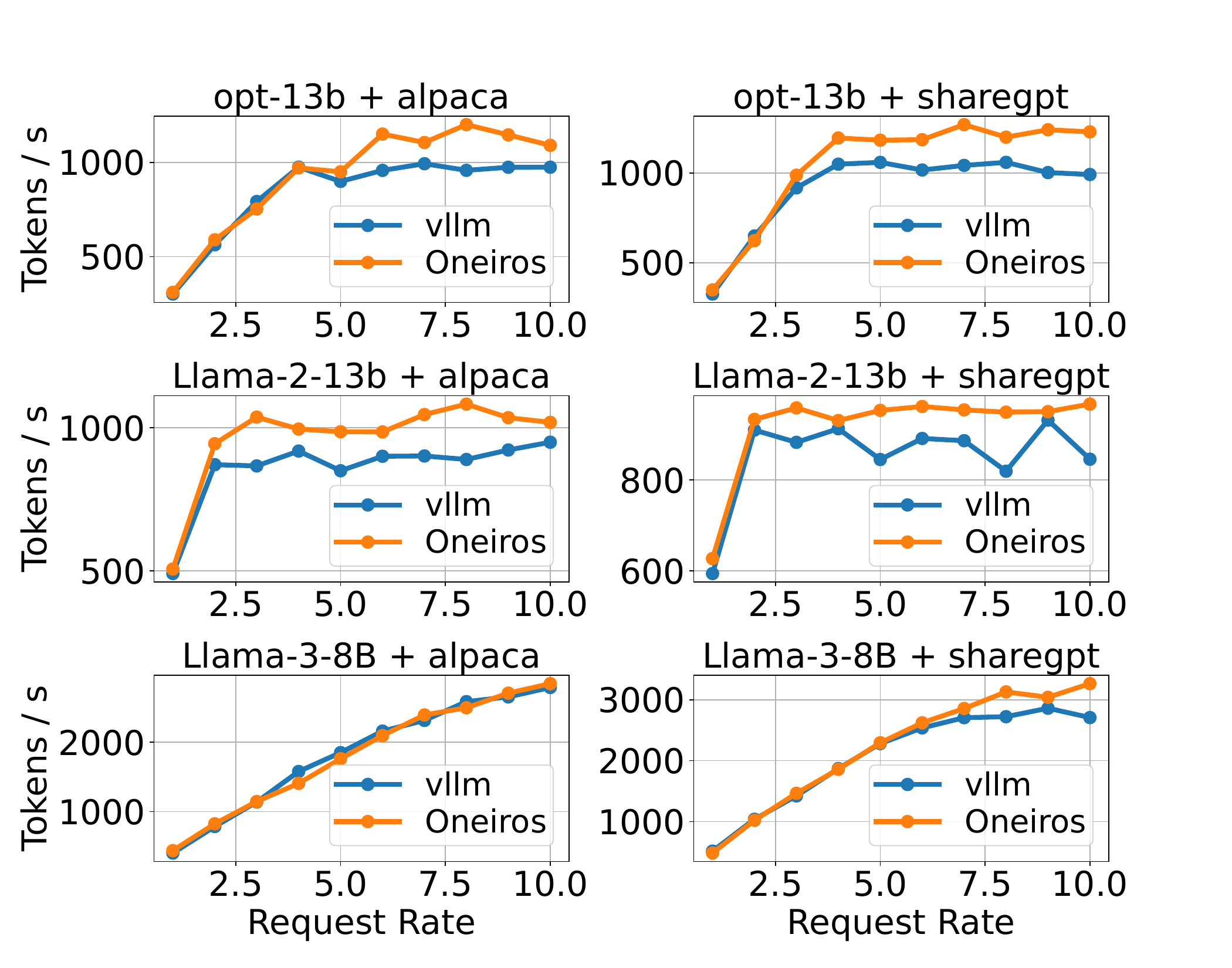}
        \vspace{-5mm}
		\subcaption{Throughput.}
		\label{fig_spatial_mig_3}
	\end{minipage}
    \vspace{-4mm}
	\caption{Comparison between vllm and \LLMPipe under non-strict physical isolation. \LLMPipe achieves slightly higher throughput, but reduces tail TBT and TTFT latency significantly.
    }
	\label{fig_spatial_mig}
\end{figure*}

\noindent{\textbf{Strict Physical Isolation:}}
\LLMPipe also maintains strong performance under strict physical isolation.
As shown in Figure~\ref{fig_spatial_mig}, it significantly reduces both tail TBT and TTFT latency as arrival rates increase, achieving an average 57.4\% reduction in P99 TBT, 34.8\% reduction in P99 TTFT, and a 7.9\% increase in throughput across different models and load conditions.
Notably, strict physical isolation between models can be viewed as a special case of single-model LLM inference.
This further demonstrates the flexibility and robustness of \LLMPipe across diverse deployment scenarios.

%% file: 07_04.tex
\begin{figure}[t]
    \centerline{\includegraphics[width=\columnwidth, trim = 32mm 2mm 50mm 4mm, page=1, clip=true]{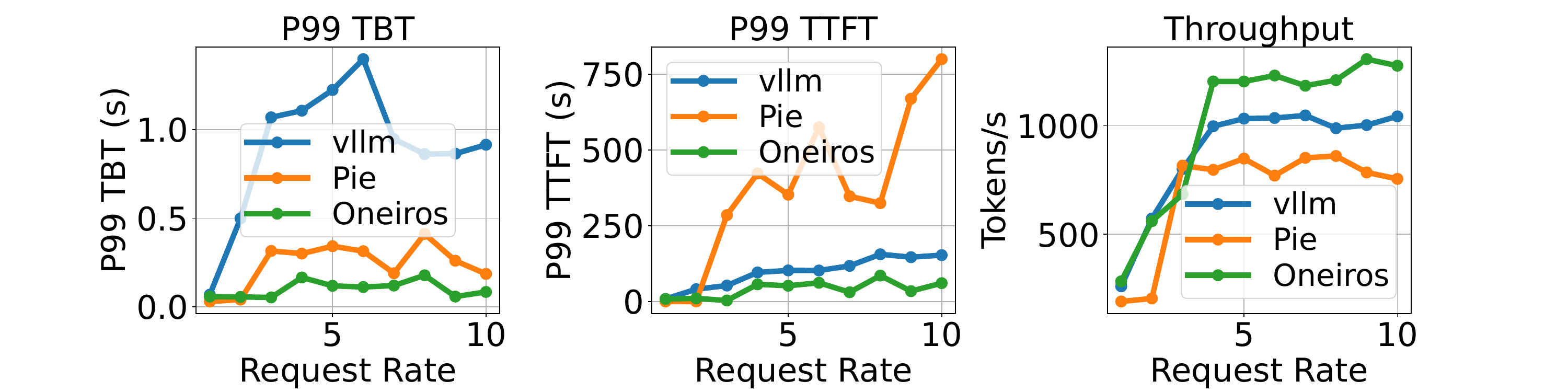}}
    \vspace{-4mm}
	\caption{
    \LLMPipe reduces tail TBT and TTFT latency while delivering higher throughput than Pie.
    }
	\label{fig_pie}
\end{figure}

\vspace{-2mm}
\subsection{\LLMPipe vs KV Cache Swapping}
\label{section_results_vs_pie}
Compared to KV-cache swapping, \LLMPipe achieves higher CPU-GPU bandwidth via unidirectional data transfer (\S~\ref{section_kv_swap}), improving throughput.
We evaluate \LLMPipe against Pie~\cite{xu2024pie}, a KV cache swapping based method.
Figure~\ref{fig_pie} shows the tail latency and throughput for serving  OPT-13b on the Alpaca dataset of vLLM, Pie, and \LLMPipe.
GPU memory usage follows Table~\ref{tab:model_comb}, consistent with \S~\ref{section_effective_layer} and \S~\ref{section_ablation}.
Compared to Pie, \LLMPipe reduces tail TBT and TTFT\footnote{We include the queuing delay in TTFT (TTFT = time from request arrival to generation of the first token = queuing delay + prefill time). Hence, by improving the throughput, \LLMPipe effectively reduces the queuing delay significantly, thereby reducing TTFT. } latency by 35.0\% and 93.6\%, respectively, while improving throughput by 47.1\%.
This improvement is attributed to the ability of \LLMPipe to avoid the overhead of backing up KV cache to CPU memory, thereby enabling more efficient utilization of CPU-GPU bandwidth. \par

Beyond performance gains, \LLMPipe offers greater flexibility in temporal sharing scenarios than KV-cache–swapping approaches.
In these scenarios, KV cache may be maintained only for models actively generating output tokens, while inactive models may not retain any.
As a result, KV cache swapping based methodologies cannot reclaim memory from inactive models, limiting the ability to dynamically expand GPU memory for the KV cache of active models.
\LLMPipe overcomes this limitation by repurposing the memory allocated to inactive models, even in the absence of KV cache. \par

%% file: 07_05.tex
\begin{figure}[t]
    \centerline{\includegraphics[width=\columnwidth, trim = 32mm 2mm 50mm 4mm, page=1, clip=true]{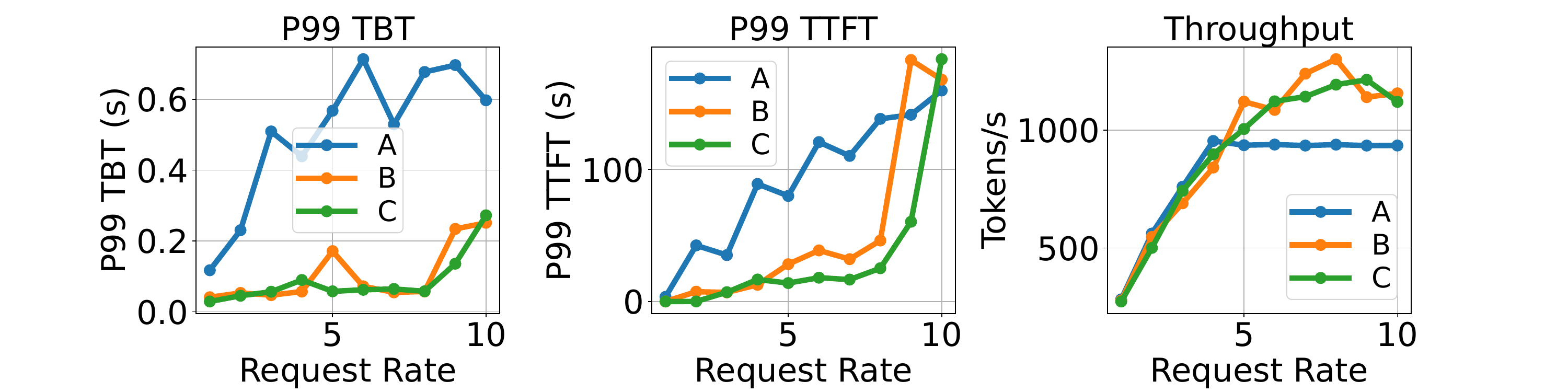}}
    \vspace{-4mm}
	\caption{Using $m=\alpha+2$ (B/C) yields lower latency and higher throughput than consistently using $m = \alpha + 1$ (A).
    }
	\label{fig_layer_selection}
\end{figure}

\subsection{Effectiveness of Our Layer Selection Strategy}
\label{section_effective_layer}

The layer selection strategy is important for enabling \LLMPipe to effectively pipeline GPU computation with CPU-GPU parameter transfer, thereby minimizing performance overhead.
The main trade-off involves selecting the number of layers ($m$, as defined in \S~\ref{section_design_remapping_which_layer}) to transfer from CPU to GPU when $\alpha$ layers are remapped.
Using $m=\alpha+1$ reduces the amount of data transfer, while $m=\alpha+2$ better addresses data dependencies by allowing full pipelining through a double-buffering mechanism.
We present an empirical evaluation of three configurations:
(A) using $m=\alpha+1$ consistently,
(B) using $m=\alpha+2$ consistently, and
(C) a dynamic scheme that selects $m=\alpha+1$ for small $\alpha$ and switches to $m=\alpha+2$ as $\alpha$ increases.
Figure~\ref{fig_layer_selection} presents the tail latency and throughput of serving the OPT-13b model on the Alpaca dataset using the different layer selection strategies described above.
Our findings show that using $m=\alpha+2$ -- where transferred layers share a two-layer memory region to enable double buffering -- is necessary for optimal performance.
Using $m=\alpha+1$ consistently (A) can result in a reduction in $12.7\%$ throughput compared to using $m=\alpha+2$ consistently (B) or switching to $m=\alpha+2$ as $\alpha$ increases (C).
The reduction in the amount of data transfer when using  $m=\alpha+1$ does not outweigh the overhead caused by disruptions to the GPU execution pipeline. \par

%% file: 07_06.tex
\subsection{When and How Many Layers to Remap?}
\label{section_ablation}
We conduct ablation studies to explore additional design parameter choices in \LLMPipe.
While parameter remapping needs to be initiated only when KV cache capacity is exhausted, determining whether to ``halt'' remapping during off-peak periods is also important (\S~\ref{section_results_free}).
Moreover, more aggressive remapping can help avoid recomputation and reduce tail latency, but it may increase average latency due to increased data transfer overhead (\S~\ref{section_results_trade_off}).

\begin{figure}[t]
    \centerline{\includegraphics[width=\columnwidth, trim = 8mm 8mm 8mm 8mm, page=1, clip=true]{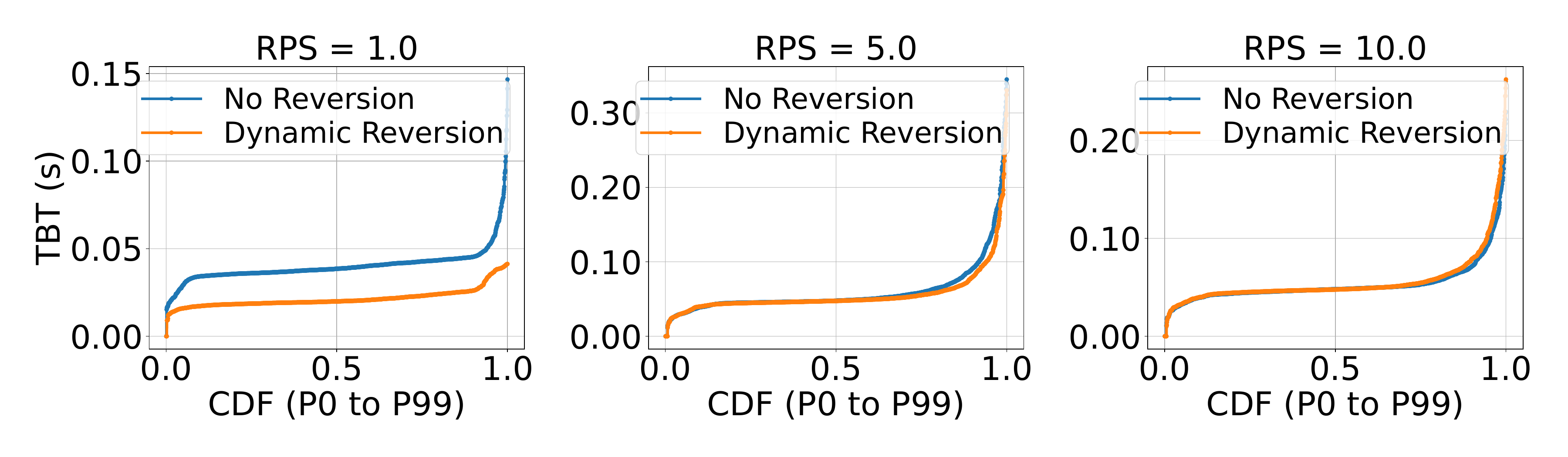}}
    \vspace{-4mm}
	\caption{TBT CDF comparison between without and with \textit{Dynamic Reversion} for OPT-13b+Alpaca. \textit{Dynamic Reversion} effectively reduces TBT latency for non-peak periods. }
	\label{fig_single_model_free_cdf}
\end{figure}

\vspace{-1mm}
\subsubsection{When to halt remap?}
\label{section_results_free}
\LLMPipe monitors KV cache utilization at runtime and detects when free KV cache is available, typically during non-peak periods. 
For example, real-world traces show that request volume during non-peak times drops to approximately 20\% of that seen during peak periods~\cite{stojkovic2024dynamollm}.
To optimize performance in such periods, \LLMPipe supports ``halting'' parameter remapping by reconfiguring the previously remapped memory space back for parameter usage. 
We name this mechanism \textit{Dynamic Reversion} and evaluate it in this section. \par

Figure~\ref{fig_single_model_free_cdf} compares the CDF of TBT latency for OPT-13B on the Alpaca dataset, with and without \textit{Dynamic Reversion}.
At peak load (10.0 RPS), \textit{Dynamic Reversion} offers limited benefit since remapping is consistently required to avoid recomputation.
During non-peak periods (1.0 RPS), however, it improves performance substantially—reducing P50 latency by 49\% and consistently lowering TBT.

\begin{figure}[t]
    \centerline{\includegraphics[width=\columnwidth, trim = 8mm 8mm 8mm 8mm, page=1, clip=true]{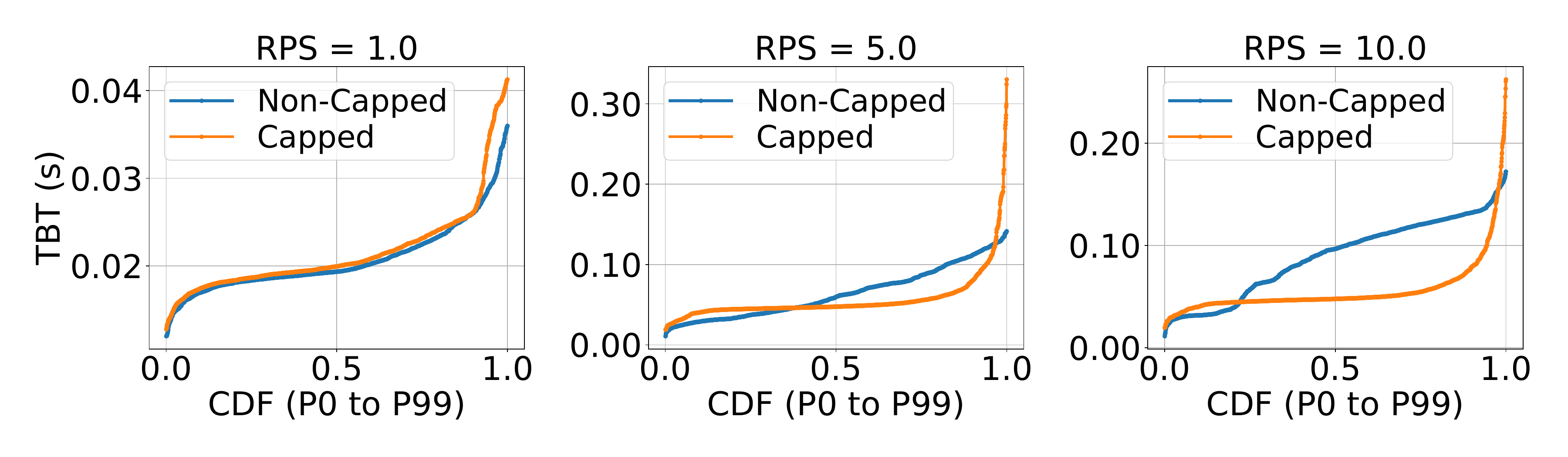}}
    \vspace{-4mm}
	\caption{Capped vs non-capped \textit{remapping percentage}. To prevent parameter loading from becoming the bottleneck, the \textit{remapping percentage} must be appropriately capped.  }
	\label{fig_tradeoff}
\end{figure}

\vspace{-1mm}
\subsubsection{How many layers to remap?}
\label{section_results_trade_off}
A key design parameter in \LLMPipe is the \textit{remapping percentage} -- the proportion of model parameter memory dynamically repurposed for KV cache.
Higher percentages help sustain peak request loads that are hard to predict offline, but excessive remapping increases CPU–GPU transfer overhead, degrading concurrent request performance.
We evaluate the impact of remapping percentage using OPT-13b with Alpaca dataset.
As shown in Figure~\ref{fig_tradeoff}, aggressive remapping reduces tail latency under light workloads (e.g., 1.0 RPS), but increases average latency due to higher CPU-GPU data transfer overhead.
Capping the remapping percentage yields a better balance—maintaining low average latency across workloads while still reducing tail latency.
At 10 RPS, the capped strategy reduces P99 latency by 58\% and P50 by 44\%.

%% file: 08_discussion.tex
\noindent{\textbf{Compatibility with other LLM Optimization Solutions:}} 
\LLMPipe integrates seamlessly with state-of-the-art LLM inference optimizations.
When the \textit{Dynamic Remapping Engine} reaches its remapping percentage limit, it can leverage CPU offloading methods such as NEO~\cite{jiang2024neo} and FlexInfer~\cite{naflexinfer} to maintain service quality.
Unlike KV-cache swapping, \LLMPipe requires no CUDA kernel modifications, ensuring compatibility with advanced kernels like Pod-Attention~\cite{kamath2024pod} and LServe~\cite{yang2025lserve}.
As a scheduler-independent memory engine, \LLMPipe supports temporal~\cite{bai2020pipeswitch, gujarati2020serving, xiao2020antman, fu2024serverlessllm, patke2024queue}, spatial~\cite{romero2021infaas, strati2024orion, dhakal2020gslice, choi2022serving, xu2022igniter, han2024kace, wang2024improving}, and hybrid~\cite{duan2024muxserve, yu2025prism} sharing policies.
By activating parameter remapping only when GPU memory is insufficient for KV cache, \LLMPipe serves as a fallback for schedulers that mispredict memory demand under dynamic workloads. \par

\noindent{\textbf{Evolving LLM Trends:}} The growing popularity of multi-agent workflows~\cite{talebirad2023multi, chan2023chateval, li2024more, han2024llm, kim2025cost} makes \LLMPipe well-suited for scenarios involving temporal GPU resource sharing. 
Furthermore, as the trend toward deploying multiple smaller-scale LLMs on a single host becomes more prevalent~\cite{wang2024comprehensive}, \LLMPipe offers additional benefits by enabling efficient memory reuse across co-located models. \par

\noindent{\textbf{Evolving Hardware Trends:}}
The Grace Hopper Superchip highlights a clear trend toward higher CPU-GPU bandwidth in future hardware.
Its successor, GB200~\cite{GB200}, preserves the 900 GB/s bidirectional bandwidth while scaling to 2 GPUs per node and up to 72 GPUs in an NVLink-connected rack-scale system. 
This architecture ensures that \LLMPipe remains compatible with multi-GPU and distributed LLM inference optimizations~\cite{sun2024llumnix, fu2024serverlessllm, yu2025prism}. 
The improvements in CPU DRAM bandwidth, further reinforces this trend.
Unlike traditional DDR memory, the LPDDRX memory used in GH200 offers significantly higher bandwidth with lower power consumption, enabling such high CPU-GPU transfer rates~\cite{lee2018leveraging, park2024lpddr, li2018performance}. \par

\noindent{\textbf{Generic Applicability of \LLMPipe across GPU Types:}}
\LLMPipe can also be deployed on earlier hardware generations, such as PCIe-based GPU systems with lower CPU-GPU bandwidth. 
The \Engine in \LLMPipe dynamically determines the number of layers to remap based on the measured CPU-GPU transfer time per layer. 
On lower-bandwidth systems, \LLMPipe remains effective by adaptively remapping fewer layers (\S~\ref{section_design_remapping_how}). 
In such environments, it can be further improved by leveraging CPU offloading (as discussed in \S~\ref{section_compute_swap}) to achieve additional performance gains. 
\par

\noindent{\textbf{Other Related Work:}}
Although prior work has explored using CPU memory as an extension of GPU memory, the focus has predominantly been on training scenarios, such as DeepSpeed~\cite{rajbhandari2020zero, yao2024training, ren2021zero, rajbhandari2021zero} and others~\cite{huang2020swapadvisor, wang2022melon, yuan2024accelerating, lim2021zico}.
Inference serving is latency-sensitive rather than throughput-oriented, making memory offloading particularly challenging.
FlexGen~\cite{sheng2023flexgen} performs on-demand CPU–GPU data migration, often delaying computation as transfers complete~\cite{xu2024pie}.
Pie~\cite{xu2024pie} was the first to exploit higher CPU–GPU bandwidth via KV-cache swapping, but its performance suffers from bidirectional transfer overhead.
\LLMPipe differs fundamentally—serving as the first system to enable efficient \textit{multi-tenant} LLM inference with dynamic memory remapping.

%% file: 09_conclusion.tex
We presented \LLMPipe, a Dynamic Remapping Engine that elastically repurposes model parameter memory to expand KV-cache capacity for LLM inference, achieving low tail latency and high throughput. \LLMPipe exploits modern CPU–GPU bandwidth to overlap parameter transfers with computation, minimizing remapping overhead. It integrates seamlessly with existing serving systems and supports diverse 
GPU sharing and scheduling policies.

\noindent{\textbf{Acknowledgement:}}
We thank the anonymous reviewers for their valuable feedback and Dimitrios Liakopoulos for proofreading.
This research was supported in part by NSF grant numbers \#2326894 and \#2425655, the UT ECE junior faculty start-up fund, UT iMAGiNE consortium, an award from the UT Machine Learning Lab (MLL), the AMD Chair Endowment, the Amazon Research Award, and Vista GPU Cluster through CGAI and TACC at UT Austin.